\documentclass[aps,prd,amsmath,amssymb,reprint,superscriptaddress]{revtex4-2}
\usepackage{graphicx}
\usepackage{amsmath}
\usepackage{dcolumn}
\usepackage{bm}
\usepackage{bbold}
\usepackage{color}
\usepackage{amsfonts}
\usepackage{amssymb}
\usepackage{mathrsfs}
\usepackage{braket}

\usepackage{caption}
\usepackage{subcaption}

\begin{document}  

\title{Rare Events Detected with a Bulk Acoustic Wave High Frequency Gravitational Wave Antenna}

\author{Maxim Goryachev}
\affiliation{ARC Centre of Excellence for Engineered Quantum Systems, ARC Centre of Excellence for Dark Matter Particle Physics, Department of Physics, University of Western Australia, 35 Stirling Highway, Crawley WA 6009, Australia}

\author{William M. Campbell}
\affiliation{ARC Centre of Excellence for Engineered Quantum Systems, ARC Centre of Excellence for Dark Matter Particle Physics, Department of Physics, University of Western Australia, 35 Stirling Highway, Crawley WA 6009, Australia}

\author{Ik Siong Heng}
\affiliation{SUPA, School of Physics and Astronomy, University of Glasgow, Glasgow G12 8QQ, UK}

\author{Serge Galliou}
\affiliation{Department of Time and Frequency, FEMTO-ST Institute, ENSMM, 26 Chemin de l{'}\'Epitaphe, 25000, Besan\c con, France}

\author{Eugene N. Ivanov}
\affiliation{ARC Centre of Excellence for Engineered Quantum Systems, ARC Centre of Excellence for Dark Matter Particle Physics, Department of Physics, University of Western Australia, 35 Stirling Highway, Crawley WA 6009, Australia}

\author{Michael E. Tobar}
\email{michael.tobar@uwa.edu.au}
\affiliation{ARC Centre of Excellence for Engineered Quantum Systems, ARC Centre of Excellence for Dark Matter Particle Physics, Department of Physics, University of Western Australia, 35 Stirling Highway, Crawley WA 6009, Australia}

\date{\today}


\begin{abstract}

This work describes the operation of a High Frequency Gravitational Wave detector based on a cryogenic Bulk Acoustic Wave (BAW) cavity and reports observation of rare events during 153 days of operation over two separate experimental runs (Run 1 and Run 2). In both Run 1 and Run 2 two modes were simultaneously monitored. Across both runs, the 3rd overtone of the fast shear mode (3B) operating at 5.506 MHz was monitored, while in Run 1 the second mode was chosen to be the 5th OT of the slow shear mode (5C) operating at 8.392 MHz. However, in Run 2 the second mode was selected to be closer in frequency to the first mode, and chosen to be the 3rd overtone of the slow shear mode (3C) operating at 4.993 MHz. Two strong events were observed as transients responding to energy deposition within acoustic modes of the cavity. The first event occurred during Run 1 on the 12/05/2019 (UTC), and was observed in the 5.506 MHz mode, while the second mode at 8.392 MHz observed no event. During Run 2, a second event occurred on the 27/11/2019(UTC) and was observed by both modes. Timing of the events were checked against available environmental observations as well as data from other detectors. Various possibilities explaining the origins of the events are discussed. 

\end{abstract}

\maketitle


\section*{Introduction}
 
Gravitational Wave (GW) astronomy became a reality on the 14 September 2015 \cite{Collaboration:2016aa}. Since then GW interferometric detectors have provided additional channels for studying the Universe, and have complemented observations conducted with telescopes. However, current operating GW interferometric detectors are only capable of probing space time in a relatively narrow band of frequency (100-1kHz), unlike the electromagnetic spectrum, where observations may be conducted over a vast frequency range (from low RF to X-ray). Recently, High Frequency Gravitational Waves (HFGWs) have been considered as a probe for new physics,  as outlined in the recent white paper on this topic \cite{HFGW}, thus the need for GW detectors with higher frequency capabilities is well motivated and has been considered in a significant way by the community \cite{Akutsu2008,Nishizawa2008,Cruise_2012,Chou2017,Caprini_2018,Martinez_2020,Ito:2020kf,Herman20,Fujita20,Domcke21}. In this work we present the first fast signal analysis of a HFGW detector based on a high frequency acoustic cavity, with plans of further analysis of slow signals/fluctuations to follow in a subsequent work.  
\begin{figure}
\centering
            \includegraphics[width=0.5\textwidth]{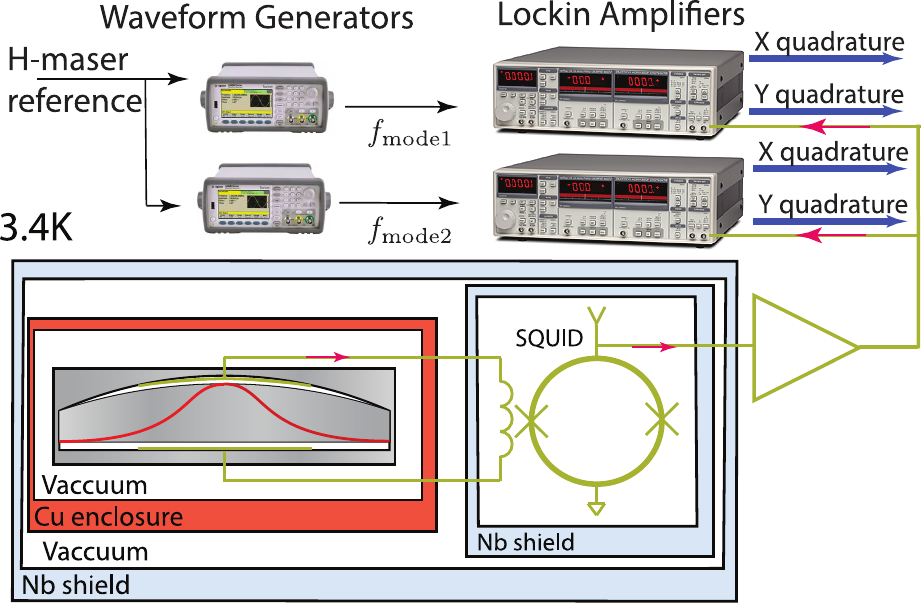}
            
    \caption{Experimental setup showing the BAW cavity connected to a SQUID amplifier and shielding arrangement. Note that 4K, 50K shields as well as the stainless still vacuum chamber are not shown. }%
   \label{setupSIG}
\end{figure}

\begin{figure}
\centering
            \includegraphics[width=0.5\textwidth]{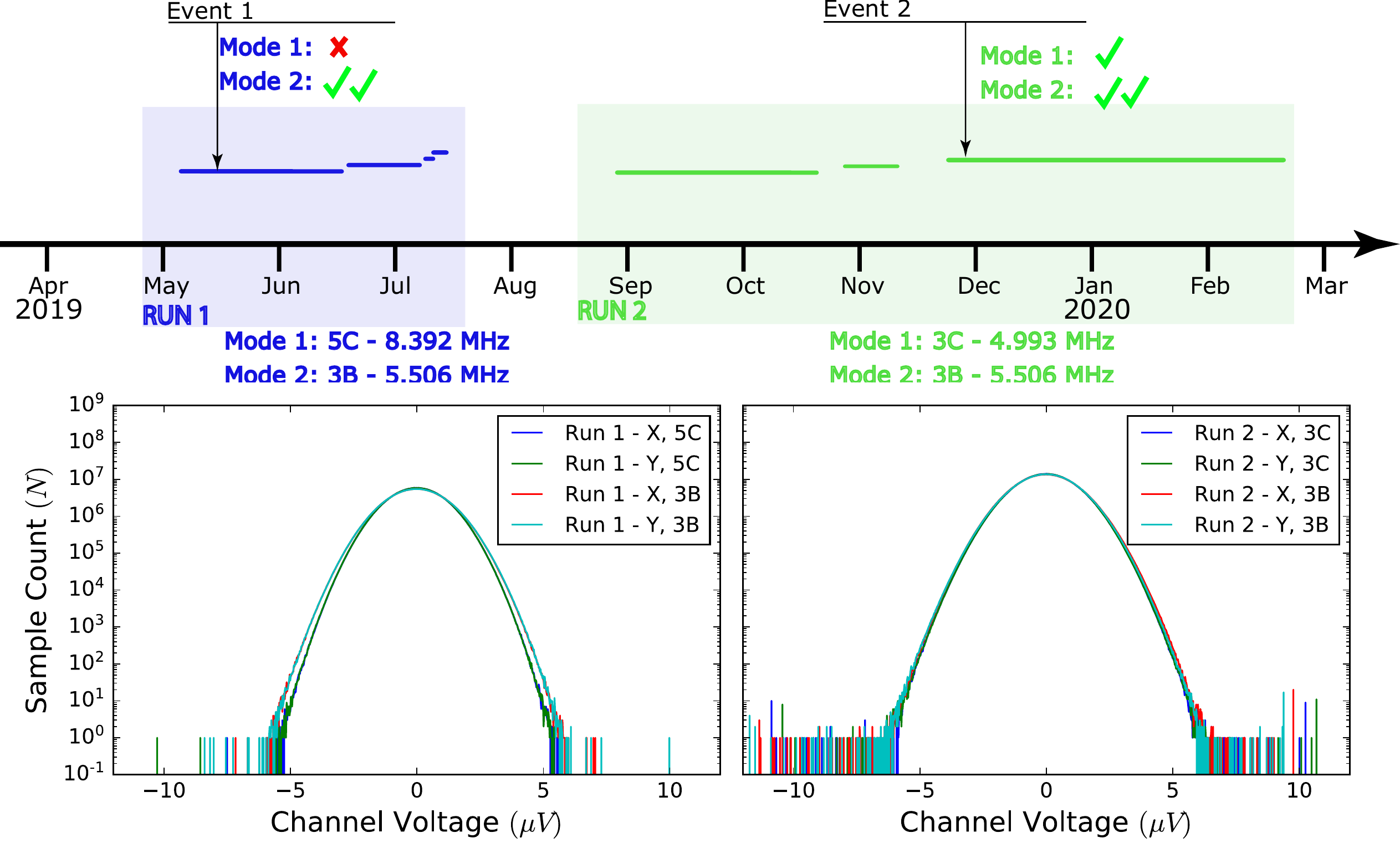}

    \caption{Time-line of the described experiment as well as histogram of the total data collection at the detector output. Blue and green lines on the time-line show separate data acquisition periods for two runs. Arrows point dates of the two observed events.}%
   \label{timeline}
\end{figure}

\begin{figure}
     \centering
            \includegraphics[width=0.5\textwidth]{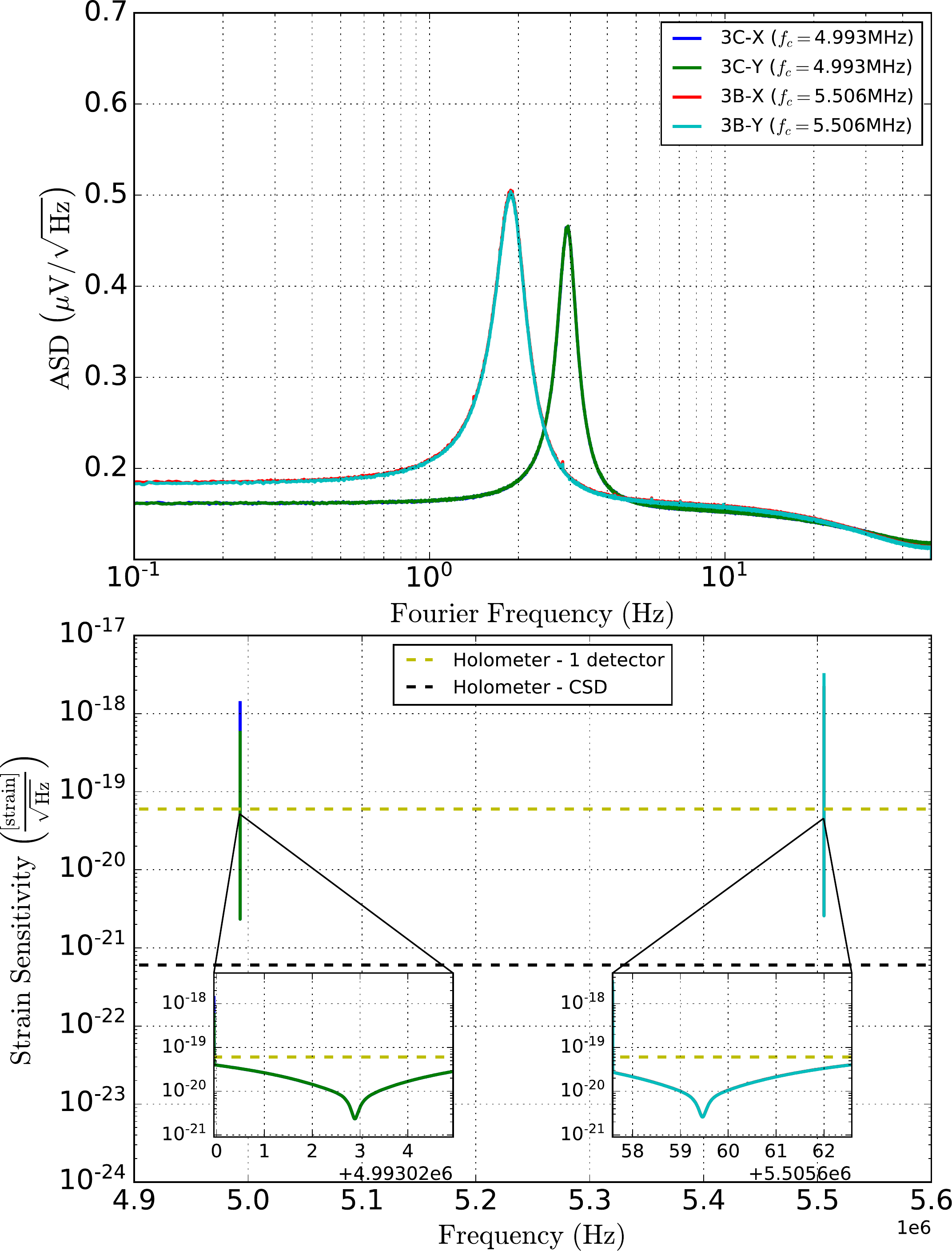}

    \caption{The top figure displays the averaged amplitude spectral density of each output channel of the lock-ins for the longest continuous data taking run, here each mode has been demodulated from the carrier. The bottom figure shows the corresponding spectral strain sensitivity determined for each trace, as well as the current best sensitivity in the region given by the Holometer experiment \cite{Chou2017}, which uses the cross spectral density (CSD) of two identical interferometers to search for HFGWs.}%
   \label{psd}
\end{figure}

\begin{figure}
\centering
            \includegraphics[width=0.5\textwidth]{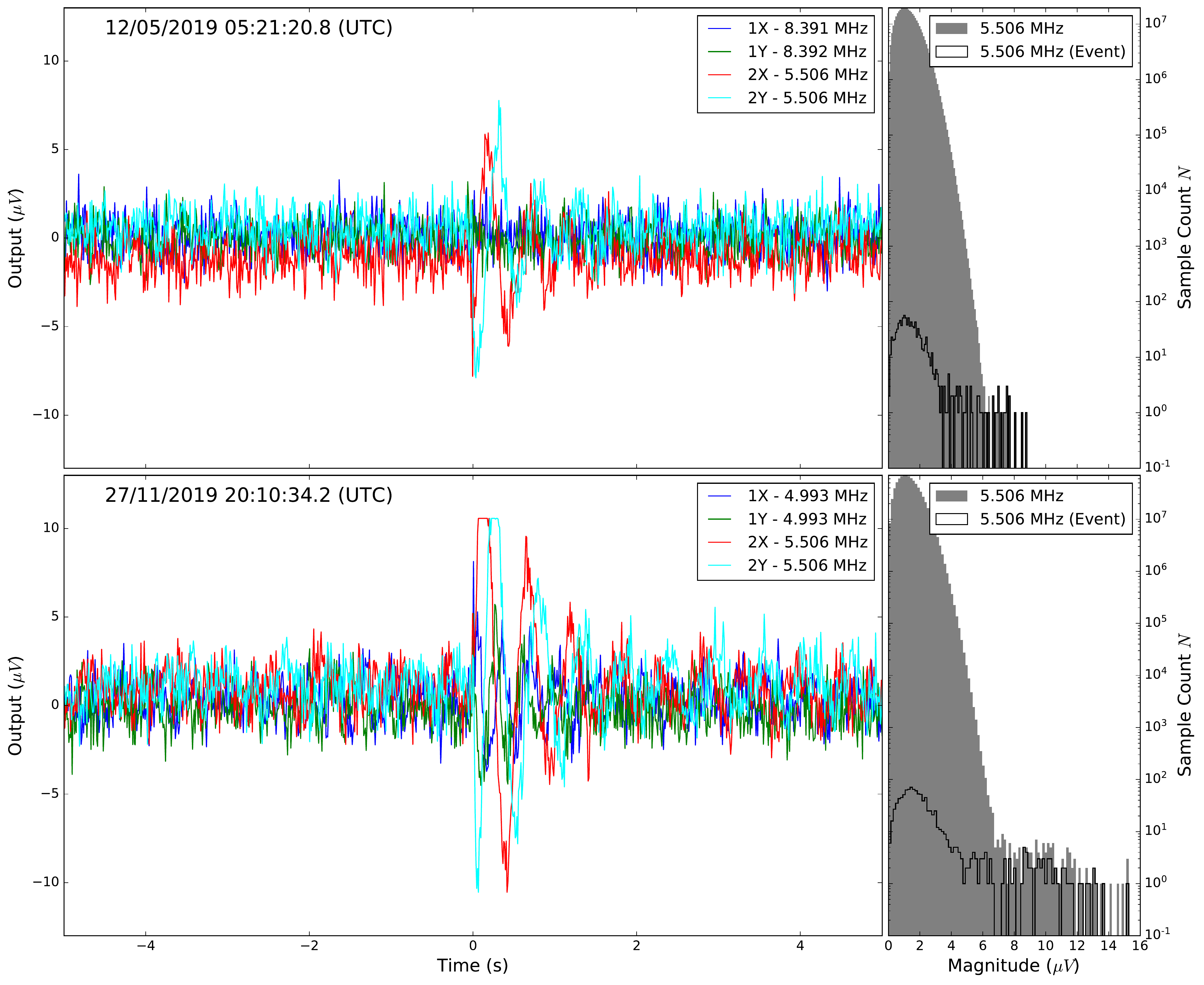}

    \caption{Time series traces for the two event signals detected by the system. Each plot shows two quadratures for each mode. Also shown are histograms of the output magnitude samples from the 3B mode only from both the entire corresponding run (grey) and just the 10s of data around the event (black). It is clear from this plot that the overwhelming majority of non-Gaussian outliers are due to these signals.}%
   \label{events}
\end{figure}

The HFGW detector employed in this work was proposed in 2014 \cite{Goryachev:2014ab}, and is based on the principles of the resonant-mass GW detector, which were operational from the 1990s mainly as resonant-bars or spheres \cite{Astone2003,PhysRevLett.74.1908,PhysRevD.54.1264,WWJ93,Harry1996,Baggio2005,Aguiar_2008}. Besides our detector, different variants of the macroscopic resonant-mass detector have been recently proposed to detect HFGWs \cite{PhysRevLett.110.071105,Aggarwal2020}. In the current form, our system is based on extremely high quality factor quartz Bulk Acoustic Wave (BAW) cavities \cite{ScRep,apl2} and a Superconducting Quantum Interference Device (SQUID) amplifier operating at 3.4K in a closed cycle cryocooler, with a schematic shown in Fig~\ref{setupSIG}. This detector is essentially a multi-mode resonant-mass GW antenna working on different overtones (OTs) of its acoustic modes that are sensitive to GWs. The high quality of these modes (reaching $4\times 10^9$ for high OT modes \cite{ScRep}) has been achieved by employing energy trapping technology of the quartz crystal plate. Due to the piezoelectric properties of quartz crystals, the systems acoustic modes which display sensitivity to GW strains can be read out via piezoelectric coupling to capacitive electrodes held at small gaps from the vibrating body. Three different acoustic mode families in the bulk of the device can be observed with this setup, longitudinal phonon or A modes, quasi fast shear or B modes, and quasi slow shear or C modes. The shear wave spitting arises from anisotropies in the crystal creating different phonon speeds depending on the wave's polarisation.\\
\\
The resonating device is composed of an approximately 1mm thick plano-covex BVA SC-cut \cite{Tiers1, 1537081, pz:1988zr} quartz plate 30mm in diameter, situated in a copper enclosure with only two isolated signal pins protruding  \cite{Galliou2013}. This enclosure is placed under a dedicated vacuum held at pressures lower than $10^{-6}$ mbar, and is isolated from the vacuum of the cryocooler chamber. The readout electrodes are inductively coupled to a Magnicon SQUID sensor of input inductance 400 $n$H, which when appropriately biased provides linear amplification and effective current-to-voltage transduction characterised by a transimpedance of $1.2~\textrm{M}\Omega$. This readout system allows for an extremely low noise floor set by the magnetic flux noise of the SQUID, which was inferred from the output voltage noise floor to be $\approx 1.1~\mu\phi_0/\sqrt{\textrm{Hz}}$ at 5.506 MHz,where $\phi_0$ is the flux quantum. The setup has been demonstrated previously, which confirmed the low noise operation limited only by the fundamental Nyquist thermal fluctuations due to ambient temperature, combined with the intrinsic SQUID amplifier readout noise \cite{ourSQUID}. To improve thermal and electrical isolation, both the quartz BAW cavity and the SQUID were contained in a large niobium shield, which was also used to support the structure. The whole setup was also covered by 4K and 50K anti-radiation shields as well as a vacuum chamber ensuring the pressure of $\sim 3\times10^{-6}$ mbar. The output of the SQUID system was further amplified at room temperature, and the signal was split between two standalone SRS SR844 lock-in amplifiers. Each of these lock-ins were tuned to near a particular resonance frequency of the BAW cavity to down convert its signal close to DC. The corresponding frequency reference signals were produced by commercial waveform signal generators locked to a Hydrogen maser. In this arrangement, each lock-in amplifier produced two quadratures of a signal giving four available output channels in total. The resulting signals were digitized using a multichannel acquisition system with the sampling rate of 100 Hz. The setup was located in a basement laboratory in Perth, Western Australia ($31.98^\circ$ S, $115.819^\circ$ E).\\
The time-line of the HFGW search is shown in Fig.~\ref{timeline}. Operation was split between two runs due to a data acquisition system upgrade, the first observational run saw 1616.7 hours of active operation at a duty rate of $98.27\%$, with the following run active for 4055.6 hours at a rate of $91.89\%$. During the first run the 5th OT of the slow shear mode at $8.392$~MHz (5B) and the 3rd OT of the fast shear mode at $5.506$~MHz (3B) of the BAW cavity were continuously monitored. For the second run, the 3rd OT of the slow shear mode at $4.993$~MHz (3C) and the 3B mode were monitored. The modes available for monitoring at this stage of the experiment were limited by the SQUID electronics having a 3dB bandwidth of only 2.1MHz. This limited our choice of overtone modes to those under 20.4 MHz with typical quality factors in the tens of millions. This is a minor technical obstacle that will be overcome with future upgrades to the setup, allowing for higher OT modes with better quality factors to be monitored. For the 3B,3C and 5C modes quality factors where previously reported to be 44, 48 and 10.7 million respectfully, as detailed in Ref. \cite{ourSQUID}. For consistency we estimated the Quality factor by fitting to the power spectrum of the current data sets, this showed close agreement to the previously reported values.\\
\\
Fig.~\ref{psd} presents the Amplitude Spectral Density of both quadratures demodulated near two resonance frequencies for the longest continuous data acquisition (2227.8 hours). All signals demonstrate clear Lorentzian peaks corresponding to thermal (Nyqvist) noise of the acoustic modes of the BAW device  \cite{ourSQUID}, while broadband features are set by the flux noise floor of the SQUID. We have also included the spectral strain sensitivity of the device which can be calculated from the corresponding voltage PSDs at the SQUID output using the transfer function:
\begin{equation}
H(i\omega)=\frac{\kappa_\lambda^2 \omega_\lambda^2 Z_\textrm{SQUID}^2\frac{-\omega^2}{2}h_0\xi}{(i\omega)^2+\tau_\lambda^{-1}i\omega+\omega_\lambda^2}
\end{equation}
Where $Z_\textrm{SQUID}$ is the SQUID transimpedance, $\kappa_\lambda$ is an experimentally found electromechanical coupling constant that relates charge on the BAW electrodes to displacement of the crystal \cite{Goryachev:2014ab}, $\omega_\lambda$ is the mode frequency with mode decay time $\tau_\lambda$, $h_0$ is the crystal thickness and $\xi$ is the weighting matrix term which parametrises the coupling of the BAW to gravitational waves, it depends heavily on the effectiveness of phonon trapping in the acoustic modes.\\
\\ 
Comparing our system to the only other HFGW detector in this frequency range; Fermi Lab's Holometer interferometer \cite{Chou2017}, which provided spectral strain measurements from 1-13 MHz with 130 hours of data collection, we see that we are within two orders of magnitude of the sensitivity given by the cross spectral density of their two 39m long interferometers. With future generations of the quartz-based detector we will be able to further increase sensitivity by using modes of higher quality factor, and also explore higher frequencies without voiding the long wavelength approximation which limits larger scale detectors. It is also of note that the reported Holometer results were not sensitive to fast transient signals, such as the events we present here.\\
\\
The total observation time for Runs 1 and 2  was 3,672.3 hours or 153 days. During this observation time, only two strongly significant events were detected: one for Run 1 (12 March 2019 at 13:21:20.8 AWST) and one for Run 2 (28 November 2019 at 04:10:34.2 AWST). Signal traces of these events are shown in Fig.~\ref{events} where each plot shows all four data channels (two quadratures of the two modes). Event 1 happens 5503 minutes after the start of the corresponding data acquisition, whereas Event 2 happens 2247 minutes after, no other acquisition periods displayed events in these time ranges. Each event was a transient ring down with a measured decay constant of $\approx$ 1 - 2 seconds, consistent with the known quality factors of the BAW cavity modes. Thus, these events were most likely to have originated from within the BAW cavity and not any other part of the detection chain. The interaction itself would be well described by a single short pulse energy deposit in an acoustic mode. By calculating the kinetic energy of the resonating crystal, we estimate the energy deposition of events 1 and 2 to be of the order of tens of $m$eV.\\ 
\\
It can be noted that Event 1  was visible only for the 3B mode while the 5C mode stays unperturbed at this level of sensitivity. The frequency difference between these modes  was approximately $2.886$~MHz. For the event 2, the strongest signal was again produced for the 3B mode while still visible on the 3C modes. These modes were separated only by $513$~kHz. In the second case, the modes were not only closer in frequency, but also of the same mode order (three variations of the acoustic field in the thickness). This might suggest that the experiment displays either frequency or mode order sensitivity to these events, in the process of detection. Additionally mode polarization may also have some effect on the signal shape in certain modes, as slow and fast shear modes are almost (but not completely) mutually orthogonal with respect to the orientation of mechanical displacement relative to the direction of wave propagation.\\
\\

We have compared timing of the observed events against other known observations of various nature available to the authors:\\
(1) \textit{weather perturbations} (thunder in particular) may cause various disturbances to electrical circuits. Such events were ruled out based on the fact that corresponding  time traces do not demonstrate any relation to acoustic resonances of the crystal. In any case both events correspond to very calm days: the day of Event 1 was sunny with $31^\circ$C temperature and $18$km/hour wind, the night of Event 2 was clear with $13-18^\circ$C temperatures and $21$km/hour wind; \\
(2) \textit{earthquakes} are known sources of acoustical vibration. Although, corresponding vibrational frequency were much lower than the sensitive frequency region. Moreover, the BAW cavity  was extremely well isolated from the ambient acoustic environment. No earthquakes in Australia were reported for the times of the events \cite{quake}; 
(3) \textit{LIGO/VIRGO Collaboration} reports no events on the corresponding days \cite{GraceDb};\\
(4) \textit{meteor events and cosmic showers.} To the authors knowledge no documented meteor events or cosmic showers have occurred in the experiments location during the requisite time periods.;\\
(5) \textit{fast radio bursts}. No documented fast radio burst events were detected in time near the events described in this work \cite{FRBlist};\\
(6) \textit{Acoustic Lorentz InvariancE ExperimeNt} (ALIEN) \cite{Goryachev:2018aa,Lo2016} is a sister experiment running to detect Lorentz Invariance in the matter sector using quartz BAW oscillators working at $5$~MHz in the same basement of the building (around 50m away). No signals at the same times as the observed events were detected by this experiment.

Despite assigning the observed events to the BAW cavity itself rather than to the detection system, we do not claim that they were HFGWs of any source. In fact, there is a number of physical phenomena that can produce these kinds of events ranging from internal solid state processes to highly speculative models of new physics. Here we list some of the possibilities:\\
(1) \textit{internal solid state process} and stress relaxations of the quartz plate. Although quartz resonators exhibit various complex nonlinear phenomena, they have been observed at temperatures of few mK and subject to strong excitations \cite{Goryachev:2014ac, Goryachev:2020aa}. In the current work, the detector was not connected to any electrical excitation circuit with the device terminals connected to the SQUID amplifier only. Although stress relaxation is a plausible explanation, the fact that some events have impact only on shear modes suggests that the stress is distributed in plane;\\
(2) \textit{internal radioactive events.} It is known that the behaviour of acoustic devices can be altered when they are subject to ionising radiation \cite{Friedt:2005aa, Lefevre:2009aa}. These studies concern cumulative effects of radiation exposure for various applications. To the best of authors' knowledge no individual ionising radiation events has been observed;\\
(3) \textit{cosmic rays events} have been observed and detected by the previous generation of GW bar detectors working in a frequency range of a few hundred Hz \cite{Astone:2001aa,astone2008}.  For example, predicted cosmic ray event rates for NAUTILUS go as high as 107 events per day for 44.5 GeV of energy deposited into the detector, with the highest recorded event being $87$ TeV and a range of other events spanning 0.04-57 K (where energy is given in units of Kelvin). Because the BAW detector has an $\approx 10^4$ times smaller cross-section, we would expect a reduction of events per day for a given energy of the same order when comparing to those seen in NAUTILUS. For the low energies we have observed ( $\lessapprox~20~m$eV), NAUTILUS would expect some $>10^6$ events per day, however it is highly uncertain what fraction of the incident cosmic ray's energy would actually be deposited into the quartz. In future generations of this detector Cosmic ray events of this kind could be easily identified by employing muon (cosmic shower) detectors for coincidence analysis; \\
(4) \textit{fireballs} and other meteor type events in the atmosphere \cite{Spalding:2017aa, Dudorov:2020aa}. The detector used in this work should not be sensitive to atmospheric acoustic waves as it  was shielded by two layers of vacuum. Further, excitation through the support structure  was also negligible as the vibrating part of the detector crystal  was isolated from the support by etched gaps and trapping mechanism. Finally, the sensitive frequency range  was far from typical acoustic frequencies of such events; \\
(5) \textit{HFGW sources} \cite{HFGW} were the premier target of the experiment. Comparing to GWs detected by low frequency interferometric detectors, the observed events do not  appear to represent mergers of any kind due to the lack of a chirp shape, however due to the narrowband nature of the detector a merger event may still have been sampled to produce the observed impulse decay signal. The data fits best a single energy depositing event. Solving the detectors equation of motion to match the strongest observed signal (event 2) results in a required characteristic strain amplitude of $h_c \approx 2.5 \times 10^{-16}$, if we assume the detector is excited by a transient $\tau_\textrm{GW} = 1$ ms pulse of GW radiation. Such radiation could be explained by a PBH merger of $m_\textrm{PBH} < 4\times 10^{-4}M_\odot$ (which gives a maximum frequency at inspiral of 5.5 MHz),   at a distance of $D \approx 0.01$ pc \cite{HFGW}. Additionally, rapid frequency evolution of the signal due to a short coalescence time would explain signal detection in two modes separated by $\approx 500$ kHz. This discussion of possible HFGW sources is far from exhaustive, many other models such as black hole super-radiance, and exotic compact object collisions can be tested, while analysis of low non-transient slow signals could also be achieved using existing data. Further HFGW analysis and discussions will be presented in a follow up work, including future observational runs;\\
(6) \textit{domain walls}, topological defects in dark matter, etc. Although these manifestations of dark matter are proposed to be detected with a network of magnetometers \cite{Pospelov:2013aa}, it is quite viable that similar disturbances caused by topological defects in dark matter have mechanical manifestations detectable with the considered detector. This possibility has to be analysed further; \\
(7) \textit{Weakly Interactive Massive Particles} (WIMPS) \cite{Jungman:1996aa}, which are another candidates for Dark Matter, are able to deposit energy in a form of phonons in crystals. Many WIMP detectors are, in fact, cryogenically cooled high purity crystals equipped with highly sensitive superconducting phonon detectors \cite{Angloher:2016aa,Juillard:2008aa}. \\
(8) \textit{dark matter}. Other types of dark matter including composite candidates capable of producing single events in mechanical oscillators or phonons in crystals; \\
(9) \textit{axion quark nuggets} \cite{Ge:2019aa} are claimed to be able to produce seismic and acoustic waves in planet's atmosphere \cite{nuggets1} as well as explain sum other anomalies such as Sun's corona anomaly \cite{Raza:2018aa} and DAMA/LIBRA results \cite{Zhitnitsky:2020aa} as well as others. The axion quark nugget events described in the work with the alleged detection \cite{nuggets1} can be ruled out based on the same points as for atmospheric meteor events.

In conclusion, we present observation of rare events detected with a HFGW BAW cavity detector. At this point no certain claim could be made on the origins of these events. The second implementation of this detector should rule out most of the possibilities. For this, the second generation detector will consists of two detector crystals with independent SQUID and digitizer read outs. In addition, the system will be equipped with a complementary muon detector to run a coincidence analysis with cosmic rays. 

\section*{Acknowledgements}
This research was supported by the Australian Research Council (ARC) Grant No. DP190100071, along with support from the ARC Centre of Excellence for Engineered Quantum Systems (EQUS, CE170100009) and the ARC Centre of Excellence for Dark Matter Particle Physics (CDM, CE200100008).\\
\hspace{10pt}

\section*{References}
\bibliography{biblioBAW}

\begin{thebibliography}{50}%
\makeatletter
\providecommand \@ifxundefined [1]{%
 \@ifx{#1\undefined}
}%
\providecommand \@ifnum [1]{%
 \ifnum #1\expandafter \@firstoftwo
 \else \expandafter \@secondoftwo
 \fi
}%
\providecommand \@ifx [1]{%
 \ifx #1\expandafter \@firstoftwo
 \else \expandafter \@secondoftwo
 \fi
}%
\providecommand \natexlab [1]{#1}%
\providecommand \enquote  [1]{``#1''}%
\providecommand \bibnamefont  [1]{#1}%
\providecommand \bibfnamefont [1]{#1}%
\providecommand \citenamefont [1]{#1}%
\providecommand \href@noop [0]{\@secondoftwo}%
\providecommand \href [0]{\begingroup \@sanitize@url \@href}%
\providecommand \@href[1]{\@@startlink{#1}\@@href}%
\providecommand \@@href[1]{\endgroup#1\@@endlink}%
\providecommand \@sanitize@url [0]{\catcode `\\12\catcode `\$12\catcode
  `\&12\catcode `\#12\catcode `\^12\catcode `\_12\catcode `\%12\relax}%
\providecommand \@@startlink[1]{}%
\providecommand \@@endlink[0]{}%
\providecommand \url  [0]{\begingroup\@sanitize@url \@url }%
\providecommand \@url [1]{\endgroup\@href {#1}{\urlprefix }}%
\providecommand \urlprefix  [0]{URL }%
\providecommand \Eprint [0]{\href }%
\providecommand \doibase [0]{http://dx.doi.org/}%
\providecommand \selectlanguage [0]{\@gobble}%
\providecommand \bibinfo  [0]{\@secondoftwo}%
\providecommand \bibfield  [0]{\@secondoftwo}%
\providecommand \translation [1]{[#1]}%
\providecommand \BibitemOpen [0]{}%
\providecommand \bibitemStop [0]{}%
\providecommand \bibitemNoStop [0]{.\EOS\space}%
\providecommand \EOS [0]{\spacefactor3000\relax}%
\providecommand \BibitemShut  [1]{\csname bibitem#1\endcsname}%
\let\auto@bib@innerbib\@empty
\bibitem [{\citenamefont {Abbott~et al.}(2016)}]{Collaboration:2016aa}%
  \BibitemOpen
  \bibfield  {author} {\bibinfo {author} {\bibfnamefont {B.~P.}\ \bibnamefont
  {Abbott~et al.}},\ }\href {\doibase 10.1103/PhysRevLett.116.061102}
  {\bibfield  {journal} {\bibinfo  {journal} {Physical Review Letters}\
  }\textbf {\bibinfo {volume} {116}},\ \bibinfo {pages} {061102} (\bibinfo
  {year} {2016})}\BibitemShut {NoStop}%
\bibitem [{\citenamefont {Aggarwal}\ \emph
  {et~al.}(2020{\natexlab{a}})\citenamefont {Aggarwal}, \citenamefont {Aguiar},
  \citenamefont {Bauswein}, \citenamefont {Cella}, \citenamefont {Clesse},
  \citenamefont {Cruise}, \citenamefont {Domcke}, \citenamefont {Figueroa},
  \citenamefont {Geraci}, \citenamefont {Goryachev}, \citenamefont {Grote},
  \citenamefont {Hindmarsh}, \citenamefont {Muia}, \citenamefont {Mukund},
  \citenamefont {Ottaway}, \citenamefont {Peloso}, \citenamefont {Quevedo},
  \citenamefont {Ricciardone}, \citenamefont {Steinlechner}, \citenamefont
  {Steinlechner}, \citenamefont {Sun}, \citenamefont {Tobar}, \citenamefont
  {Torrenti}, \citenamefont {Unal},\ and\ \citenamefont {White}}]{HFGW}%
  \BibitemOpen
  \bibfield  {author} {\bibinfo {author} {\bibfnamefont {N.}~\bibnamefont
  {Aggarwal}}, \bibinfo {author} {\bibfnamefont {O.~D.}\ \bibnamefont
  {Aguiar}}, \bibinfo {author} {\bibfnamefont {A.}~\bibnamefont {Bauswein}},
  \bibinfo {author} {\bibfnamefont {G.}~\bibnamefont {Cella}}, \bibinfo
  {author} {\bibfnamefont {S.}~\bibnamefont {Clesse}}, \bibinfo {author}
  {\bibfnamefont {A.~M.}\ \bibnamefont {Cruise}}, \bibinfo {author}
  {\bibfnamefont {V.}~\bibnamefont {Domcke}}, \bibinfo {author} {\bibfnamefont
  {D.~G.}\ \bibnamefont {Figueroa}}, \bibinfo {author} {\bibfnamefont
  {A.}~\bibnamefont {Geraci}}, \bibinfo {author} {\bibfnamefont
  {M.}~\bibnamefont {Goryachev}}, \bibinfo {author} {\bibfnamefont
  {H.}~\bibnamefont {Grote}}, \bibinfo {author} {\bibfnamefont
  {M.}~\bibnamefont {Hindmarsh}}, \bibinfo {author} {\bibfnamefont
  {F.}~\bibnamefont {Muia}}, \bibinfo {author} {\bibfnamefont {N.}~\bibnamefont
  {Mukund}}, \bibinfo {author} {\bibfnamefont {D.}~\bibnamefont {Ottaway}},
  \bibinfo {author} {\bibfnamefont {M.}~\bibnamefont {Peloso}}, \bibinfo
  {author} {\bibfnamefont {F.}~\bibnamefont {Quevedo}}, \bibinfo {author}
  {\bibfnamefont {A.}~\bibnamefont {Ricciardone}}, \bibinfo {author}
  {\bibfnamefont {J.}~\bibnamefont {Steinlechner}}, \bibinfo {author}
  {\bibfnamefont {S.}~\bibnamefont {Steinlechner}}, \bibinfo {author}
  {\bibfnamefont {S.}~\bibnamefont {Sun}}, \bibinfo {author} {\bibfnamefont
  {M.~E.}\ \bibnamefont {Tobar}}, \bibinfo {author} {\bibfnamefont
  {F.}~\bibnamefont {Torrenti}}, \bibinfo {author} {\bibfnamefont
  {C.}~\bibnamefont {Unal}}, \ and\ \bibinfo {author} {\bibfnamefont
  {G.}~\bibnamefont {White}},\ }\href@noop {} {\bibfield  {journal} {\bibinfo
  {journal} {arXiv:2011.12414}\ } (\bibinfo {year}
  {2020}{\natexlab{a}})}\BibitemShut {NoStop}%
\bibitem [{\citenamefont {Akutsu}\ \emph {et~al.}(2008)\citenamefont {Akutsu},
  \citenamefont {Kawamura}, \citenamefont {Nishizawa}, \citenamefont {Arai},
  \citenamefont {Yamamoto}, \citenamefont {Tatsumi}, \citenamefont {Nagano},
  \citenamefont {Nishida}, \citenamefont {Chiba}, \citenamefont {Takahashi},
  \citenamefont {Sugiyama}, \citenamefont {Fukushima}, \citenamefont
  {Yamazaki},\ and\ \citenamefont {Fujimoto}}]{Akutsu2008}%
  \BibitemOpen
  \bibfield  {author} {\bibinfo {author} {\bibfnamefont {T.}~\bibnamefont
  {Akutsu}}, \bibinfo {author} {\bibfnamefont {S.}~\bibnamefont {Kawamura}},
  \bibinfo {author} {\bibfnamefont {A.}~\bibnamefont {Nishizawa}}, \bibinfo
  {author} {\bibfnamefont {K.}~\bibnamefont {Arai}}, \bibinfo {author}
  {\bibfnamefont {K.}~\bibnamefont {Yamamoto}}, \bibinfo {author}
  {\bibfnamefont {D.}~\bibnamefont {Tatsumi}}, \bibinfo {author} {\bibfnamefont
  {S.}~\bibnamefont {Nagano}}, \bibinfo {author} {\bibfnamefont
  {E.}~\bibnamefont {Nishida}}, \bibinfo {author} {\bibfnamefont
  {T.}~\bibnamefont {Chiba}}, \bibinfo {author} {\bibfnamefont
  {R.}~\bibnamefont {Takahashi}}, \bibinfo {author} {\bibfnamefont
  {N.}~\bibnamefont {Sugiyama}}, \bibinfo {author} {\bibfnamefont
  {M.}~\bibnamefont {Fukushima}}, \bibinfo {author} {\bibfnamefont
  {T.}~\bibnamefont {Yamazaki}}, \ and\ \bibinfo {author} {\bibfnamefont
  {M.-K.}\ \bibnamefont {Fujimoto}},\ }\href {\doibase
  10.1103/PhysRevLett.101.101101} {\bibfield  {journal} {\bibinfo  {journal}
  {Phys. Rev. Lett.}\ }\textbf {\bibinfo {volume} {101}},\ \bibinfo {pages}
  {101101} (\bibinfo {year} {2008})}\BibitemShut {NoStop}%
\bibitem [{\citenamefont {Nishizawa}\ \emph {et~al.}(2008)\citenamefont
  {Nishizawa}, \citenamefont {Kawamura}, \citenamefont {Akutsu}, \citenamefont
  {Arai}, \citenamefont {Yamamoto}, \citenamefont {Tatsumi}, \citenamefont
  {Nishida}, \citenamefont {Sakagami}, \citenamefont {Chiba}, \citenamefont
  {Takahashi},\ and\ \citenamefont {Sugiyama}}]{Nishizawa2008}%
  \BibitemOpen
  \bibfield  {author} {\bibinfo {author} {\bibfnamefont {A.}~\bibnamefont
  {Nishizawa}}, \bibinfo {author} {\bibfnamefont {S.}~\bibnamefont {Kawamura}},
  \bibinfo {author} {\bibfnamefont {T.}~\bibnamefont {Akutsu}}, \bibinfo
  {author} {\bibfnamefont {K.}~\bibnamefont {Arai}}, \bibinfo {author}
  {\bibfnamefont {K.}~\bibnamefont {Yamamoto}}, \bibinfo {author}
  {\bibfnamefont {D.}~\bibnamefont {Tatsumi}}, \bibinfo {author} {\bibfnamefont
  {E.}~\bibnamefont {Nishida}}, \bibinfo {author} {\bibfnamefont {M.-a.}\
  \bibnamefont {Sakagami}}, \bibinfo {author} {\bibfnamefont {T.}~\bibnamefont
  {Chiba}}, \bibinfo {author} {\bibfnamefont {R.}~\bibnamefont {Takahashi}}, \
  and\ \bibinfo {author} {\bibfnamefont {N.}~\bibnamefont {Sugiyama}},\ }\href
  {\doibase 10.1103/PhysRevD.77.022002} {\bibfield  {journal} {\bibinfo
  {journal} {Phys. Rev. D}\ }\textbf {\bibinfo {volume} {77}},\ \bibinfo
  {pages} {022002} (\bibinfo {year} {2008})}\BibitemShut {NoStop}%
\bibitem [{\citenamefont {Cruise}(2012)}]{Cruise_2012}%
  \BibitemOpen
  \bibfield  {author} {\bibinfo {author} {\bibfnamefont {A.~M.}\ \bibnamefont
  {Cruise}},\ }\href {\doibase 10.1088/0264-9381/29/9/095003} {\bibfield
  {journal} {\bibinfo  {journal} {Classical and Quantum Gravity}\ }\textbf
  {\bibinfo {volume} {29}},\ \bibinfo {pages} {095003} (\bibinfo {year}
  {2012})}\BibitemShut {NoStop}%
\bibitem [{\citenamefont {Chou}\ \emph {et~al.}(2017)\citenamefont {Chou},
  \citenamefont {Gustafson}, \citenamefont {Hogan}, \citenamefont {Kamai},
  \citenamefont {Kwon}, \citenamefont {Lanza}, \citenamefont {Larson},
  \citenamefont {McCuller}, \citenamefont {Meyer}, \citenamefont {Richardson},
  \citenamefont {Stoughton}, \citenamefont {Tomlin},\ and\ \citenamefont
  {Weiss}}]{Chou2017}%
  \BibitemOpen
  \bibfield  {author} {\bibinfo {author} {\bibfnamefont {A.~S.}\ \bibnamefont
  {Chou}}, \bibinfo {author} {\bibfnamefont {R.}~\bibnamefont {Gustafson}},
  \bibinfo {author} {\bibfnamefont {C.}~\bibnamefont {Hogan}}, \bibinfo
  {author} {\bibfnamefont {B.}~\bibnamefont {Kamai}}, \bibinfo {author}
  {\bibfnamefont {O.}~\bibnamefont {Kwon}}, \bibinfo {author} {\bibfnamefont
  {R.}~\bibnamefont {Lanza}}, \bibinfo {author} {\bibfnamefont {S.~L.}\
  \bibnamefont {Larson}}, \bibinfo {author} {\bibfnamefont {L.}~\bibnamefont
  {McCuller}}, \bibinfo {author} {\bibfnamefont {S.~S.}\ \bibnamefont {Meyer}},
  \bibinfo {author} {\bibfnamefont {J.}~\bibnamefont {Richardson}}, \bibinfo
  {author} {\bibfnamefont {C.}~\bibnamefont {Stoughton}}, \bibinfo {author}
  {\bibfnamefont {R.}~\bibnamefont {Tomlin}}, \ and\ \bibinfo {author}
  {\bibfnamefont {R.}~\bibnamefont {Weiss}} (\bibinfo {collaboration}
  {Holometer Collaboration}),\ }\href {\doibase 10.1103/PhysRevD.95.063002}
  {\bibfield  {journal} {\bibinfo  {journal} {Phys. Rev. D}\ }\textbf {\bibinfo
  {volume} {95}},\ \bibinfo {pages} {063002} (\bibinfo {year}
  {2017})}\BibitemShut {NoStop}%
\bibitem [{\citenamefont {Caprini}\ and\ \citenamefont
  {Figueroa}(2018)}]{Caprini_2018}%
  \BibitemOpen
  \bibfield  {author} {\bibinfo {author} {\bibfnamefont {C.}~\bibnamefont
  {Caprini}}\ and\ \bibinfo {author} {\bibfnamefont {D.~G.}\ \bibnamefont
  {Figueroa}},\ }\href {\doibase 10.1088/1361-6382/aac608} {\bibfield
  {journal} {\bibinfo  {journal} {Classical and Quantum Gravity}\ }\textbf
  {\bibinfo {volume} {35}},\ \bibinfo {pages} {163001} (\bibinfo {year}
  {2018})}\BibitemShut {NoStop}%
\bibitem [{\citenamefont {Martinez}\ and\ \citenamefont
  {Kamai}(2020)}]{Martinez_2020}%
  \BibitemOpen
  \bibfield  {author} {\bibinfo {author} {\bibfnamefont {J.~G.~C.}\
  \bibnamefont {Martinez}}\ and\ \bibinfo {author} {\bibfnamefont
  {B.}~\bibnamefont {Kamai}},\ }\href {\doibase 10.1088/1361-6382/aba669}
  {\bibfield  {journal} {\bibinfo  {journal} {Classical and Quantum Gravity}\
  }\textbf {\bibinfo {volume} {37}},\ \bibinfo {pages} {205006} (\bibinfo
  {year} {2020})}\BibitemShut {NoStop}%
\bibitem [{\citenamefont {Ito}\ \emph {et~al.}(2020)\citenamefont {Ito},
  \citenamefont {Ikeda}, \citenamefont {Miuchi},\ and\ \citenamefont
  {Soda}}]{Ito:2020kf}%
  \BibitemOpen
  \bibfield  {author} {\bibinfo {author} {\bibfnamefont {A.}~\bibnamefont
  {Ito}}, \bibinfo {author} {\bibfnamefont {T.}~\bibnamefont {Ikeda}}, \bibinfo
  {author} {\bibfnamefont {K.}~\bibnamefont {Miuchi}}, \ and\ \bibinfo {author}
  {\bibfnamefont {J.}~\bibnamefont {Soda}},\ }\href {\doibase
  10.1140/epjc/s10052-020-7735-y} {\bibfield  {journal} {\bibinfo  {journal}
  {The European Physical Journal C}\ }\textbf {\bibinfo {volume} {80}},\
  \bibinfo {pages} {179} (\bibinfo {year} {2020})}\BibitemShut {NoStop}%
\bibitem [{\citenamefont {Herman}\ \emph {et~al.}(2020)\citenamefont {Herman},
  \citenamefont {F{\"u}zfa}, \citenamefont {Clesse},\ and\ \citenamefont
  {Lehoucq}}]{Herman20}%
  \BibitemOpen
  \bibfield  {author} {\bibinfo {author} {\bibfnamefont {N.}~\bibnamefont
  {Herman}}, \bibinfo {author} {\bibfnamefont {A.}~\bibnamefont {F{\"u}zfa}},
  \bibinfo {author} {\bibfnamefont {S.}~\bibnamefont {Clesse}}, \ and\ \bibinfo
  {author} {\bibfnamefont {L.}~\bibnamefont {Lehoucq}},\ }\href@noop {}
  {\bibfield  {journal} {\bibinfo  {journal} {arXiv:2012.12189 [gr-qc]}\ }
  (\bibinfo {year} {2020})}\BibitemShut {NoStop}%
\bibitem [{\citenamefont {Fujita}\ \emph {et~al.}(2020)\citenamefont {Fujita},
  \citenamefont {Kamada},\ and\ \citenamefont {Nakai}}]{Fujita20}%
  \BibitemOpen
  \bibfield  {author} {\bibinfo {author} {\bibfnamefont {T.}~\bibnamefont
  {Fujita}}, \bibinfo {author} {\bibfnamefont {K.}~\bibnamefont {Kamada}}, \
  and\ \bibinfo {author} {\bibfnamefont {Y.}~\bibnamefont {Nakai}},\ }\href
  {\doibase 10.1103/PhysRevD.102.103501} {\bibfield  {journal} {\bibinfo
  {journal} {Phys. Rev. D}\ }\textbf {\bibinfo {volume} {102}},\ \bibinfo
  {pages} {103501} (\bibinfo {year} {2020})}\BibitemShut {NoStop}%
\bibitem [{\citenamefont {Domcke}\ and\ \citenamefont
  {Garcia-Cely}(2021)}]{Domcke21}%
  \BibitemOpen
  \bibfield  {author} {\bibinfo {author} {\bibfnamefont {V.}~\bibnamefont
  {Domcke}}\ and\ \bibinfo {author} {\bibfnamefont {C.}~\bibnamefont
  {Garcia-Cely}},\ }\href {\doibase 10.1103/PhysRevLett.126.021104} {\bibfield
  {journal} {\bibinfo  {journal} {Phys. Rev. Lett.}\ }\textbf {\bibinfo
  {volume} {126}},\ \bibinfo {pages} {021104} (\bibinfo {year}
  {2021})}\BibitemShut {NoStop}%
\bibitem [{\citenamefont {Goryachev}\ and\ \citenamefont
  {Tobar}(2014)}]{Goryachev:2014ab}%
  \BibitemOpen
  \bibfield  {author} {\bibinfo {author} {\bibfnamefont {M.}~\bibnamefont
  {Goryachev}}\ and\ \bibinfo {author} {\bibfnamefont {M.~E.}\ \bibnamefont
  {Tobar}},\ }\href {\doibase 10.1103/PhysRevD.90.102005} {\bibfield  {journal}
  {\bibinfo  {journal} {Physical Review D}\ }\textbf {\bibinfo {volume} {90}},\
  \bibinfo {pages} {102005} (\bibinfo {year} {2014})}\BibitemShut {NoStop}%
\bibitem [{\citenamefont {Astone}\ \emph {et~al.}(2003)\citenamefont {Astone},
  \citenamefont {Babusci}, \citenamefont {Baggio}, \citenamefont {Bassan},
  \citenamefont {Blair}, \citenamefont {Bonaldi}, \citenamefont {Bonifazi},
  \citenamefont {Busby}, \citenamefont {Carelli}, \citenamefont {Cerdonio},
  \citenamefont {Coccia}, \citenamefont {Conti}, \citenamefont {Cosmelli},
  \citenamefont {D'Antonio}, \citenamefont {Fafone}, \citenamefont {Falferi},
  \citenamefont {Fortini}, \citenamefont {Frasca}, \citenamefont {Giordano},
  \citenamefont {Hamilton}, \citenamefont {Heng}, \citenamefont {Ivanov},
  \citenamefont {Johnson}, \citenamefont {Marini}, \citenamefont {Mauceli},
  \citenamefont {McHugh}, \citenamefont {Mezzena}, \citenamefont {Minenkov},
  \citenamefont {Modena}, \citenamefont {Modestino}, \citenamefont {Moleti},
  \citenamefont {Ortolan}, \citenamefont {Pallottino}, \citenamefont
  {Pizzella}, \citenamefont {Prodi}, \citenamefont {Quintieri}, \citenamefont
  {Rocchi}, \citenamefont {Rocco}, \citenamefont {Ronga}, \citenamefont
  {Salemi}, \citenamefont {Santostasi}, \citenamefont {Taffarello},
  \citenamefont {Terenzi}, \citenamefont {Tobar}, \citenamefont {Torrioli},
  \citenamefont {Vedovato}, \citenamefont {Vinante}, \citenamefont {Visco},
  \citenamefont {Vitale},\ and\ \citenamefont {Zendri}}]{Astone2003}%
  \BibitemOpen
  \bibfield  {author} {\bibinfo {author} {\bibfnamefont {P.}~\bibnamefont
  {Astone}}, \bibinfo {author} {\bibfnamefont {D.}~\bibnamefont {Babusci}},
  \bibinfo {author} {\bibfnamefont {L.}~\bibnamefont {Baggio}}, \bibinfo
  {author} {\bibfnamefont {M.}~\bibnamefont {Bassan}}, \bibinfo {author}
  {\bibfnamefont {D.~G.}\ \bibnamefont {Blair}}, \bibinfo {author}
  {\bibfnamefont {M.}~\bibnamefont {Bonaldi}}, \bibinfo {author} {\bibfnamefont
  {P.}~\bibnamefont {Bonifazi}}, \bibinfo {author} {\bibfnamefont
  {D.}~\bibnamefont {Busby}}, \bibinfo {author} {\bibfnamefont
  {P.}~\bibnamefont {Carelli}}, \bibinfo {author} {\bibfnamefont
  {M.}~\bibnamefont {Cerdonio}}, \bibinfo {author} {\bibfnamefont
  {E.}~\bibnamefont {Coccia}}, \bibinfo {author} {\bibfnamefont
  {L.}~\bibnamefont {Conti}}, \bibinfo {author} {\bibfnamefont
  {C.}~\bibnamefont {Cosmelli}}, \bibinfo {author} {\bibfnamefont
  {S.}~\bibnamefont {D'Antonio}}, \bibinfo {author} {\bibfnamefont
  {V.}~\bibnamefont {Fafone}}, \bibinfo {author} {\bibfnamefont
  {P.}~\bibnamefont {Falferi}}, \bibinfo {author} {\bibfnamefont
  {P.}~\bibnamefont {Fortini}}, \bibinfo {author} {\bibfnamefont
  {S.}~\bibnamefont {Frasca}}, \bibinfo {author} {\bibfnamefont
  {G.}~\bibnamefont {Giordano}}, \bibinfo {author} {\bibfnamefont {W.~O.}\
  \bibnamefont {Hamilton}}, \bibinfo {author} {\bibfnamefont {I.~S.}\
  \bibnamefont {Heng}}, \bibinfo {author} {\bibfnamefont {E.~N.}\ \bibnamefont
  {Ivanov}}, \bibinfo {author} {\bibfnamefont {W.~W.}\ \bibnamefont {Johnson}},
  \bibinfo {author} {\bibfnamefont {A.}~\bibnamefont {Marini}}, \bibinfo
  {author} {\bibfnamefont {E.}~\bibnamefont {Mauceli}}, \bibinfo {author}
  {\bibfnamefont {M.~P.}\ \bibnamefont {McHugh}}, \bibinfo {author}
  {\bibfnamefont {R.}~\bibnamefont {Mezzena}}, \bibinfo {author} {\bibfnamefont
  {Y.}~\bibnamefont {Minenkov}}, \bibinfo {author} {\bibfnamefont
  {I.}~\bibnamefont {Modena}}, \bibinfo {author} {\bibfnamefont
  {G.}~\bibnamefont {Modestino}}, \bibinfo {author} {\bibfnamefont
  {A.}~\bibnamefont {Moleti}}, \bibinfo {author} {\bibfnamefont
  {A.}~\bibnamefont {Ortolan}}, \bibinfo {author} {\bibfnamefont {G.~V.}\
  \bibnamefont {Pallottino}}, \bibinfo {author} {\bibfnamefont
  {G.}~\bibnamefont {Pizzella}}, \bibinfo {author} {\bibfnamefont {G.~A.}\
  \bibnamefont {Prodi}}, \bibinfo {author} {\bibfnamefont {L.}~\bibnamefont
  {Quintieri}}, \bibinfo {author} {\bibfnamefont {A.}~\bibnamefont {Rocchi}},
  \bibinfo {author} {\bibfnamefont {E.}~\bibnamefont {Rocco}}, \bibinfo
  {author} {\bibfnamefont {F.}~\bibnamefont {Ronga}}, \bibinfo {author}
  {\bibfnamefont {F.}~\bibnamefont {Salemi}}, \bibinfo {author} {\bibfnamefont
  {G.}~\bibnamefont {Santostasi}}, \bibinfo {author} {\bibfnamefont
  {L.}~\bibnamefont {Taffarello}}, \bibinfo {author} {\bibfnamefont
  {R.}~\bibnamefont {Terenzi}}, \bibinfo {author} {\bibfnamefont {M.~E.}\
  \bibnamefont {Tobar}}, \bibinfo {author} {\bibfnamefont {G.}~\bibnamefont
  {Torrioli}}, \bibinfo {author} {\bibfnamefont {G.}~\bibnamefont {Vedovato}},
  \bibinfo {author} {\bibfnamefont {A.}~\bibnamefont {Vinante}}, \bibinfo
  {author} {\bibfnamefont {M.}~\bibnamefont {Visco}}, \bibinfo {author}
  {\bibfnamefont {S.}~\bibnamefont {Vitale}}, \ and\ \bibinfo {author}
  {\bibfnamefont {J.~P.}\ \bibnamefont {Zendri}} (\bibinfo {collaboration}
  {International Gravitational Event Collaboration}),\ }\href {\doibase
  10.1103/PhysRevD.68.022001} {\bibfield  {journal} {\bibinfo  {journal} {Phys.
  Rev. D}\ }\textbf {\bibinfo {volume} {68}},\ \bibinfo {pages} {022001}
  (\bibinfo {year} {2003})}\BibitemShut {NoStop}%
\bibitem [{\citenamefont {Blair}\ \emph {et~al.}(1995)\citenamefont {Blair},
  \citenamefont {Ivanov}, \citenamefont {Tobar}, \citenamefont {Turner},
  \citenamefont {van Kann},\ and\ \citenamefont {Heng}}]{PhysRevLett.74.1908}%
  \BibitemOpen
  \bibfield  {author} {\bibinfo {author} {\bibfnamefont {D.~G.}\ \bibnamefont
  {Blair}}, \bibinfo {author} {\bibfnamefont {E.~N.}\ \bibnamefont {Ivanov}},
  \bibinfo {author} {\bibfnamefont {M.~E.}\ \bibnamefont {Tobar}}, \bibinfo
  {author} {\bibfnamefont {P.~J.}\ \bibnamefont {Turner}}, \bibinfo {author}
  {\bibfnamefont {F.}~\bibnamefont {van Kann}}, \ and\ \bibinfo {author}
  {\bibfnamefont {I.~S.}\ \bibnamefont {Heng}},\ }\href {\doibase
  10.1103/PhysRevLett.74.1908} {\bibfield  {journal} {\bibinfo  {journal}
  {Phys. Rev. Lett.}\ }\textbf {\bibinfo {volume} {74}},\ \bibinfo {pages}
  {1908} (\bibinfo {year} {1995})}\BibitemShut {NoStop}%
\bibitem [{\citenamefont {Mauceli}\ \emph {et~al.}(1996)\citenamefont
  {Mauceli}, \citenamefont {Geng}, \citenamefont {Hamilton}, \citenamefont
  {Johnson}, \citenamefont {Merkowitz}, \citenamefont {Morse}, \citenamefont
  {Price},\ and\ \citenamefont {Solomonson}}]{PhysRevD.54.1264}%
  \BibitemOpen
  \bibfield  {author} {\bibinfo {author} {\bibfnamefont {E.}~\bibnamefont
  {Mauceli}}, \bibinfo {author} {\bibfnamefont {Z.~K.}\ \bibnamefont {Geng}},
  \bibinfo {author} {\bibfnamefont {W.~O.}\ \bibnamefont {Hamilton}}, \bibinfo
  {author} {\bibfnamefont {W.~W.}\ \bibnamefont {Johnson}}, \bibinfo {author}
  {\bibfnamefont {S.}~\bibnamefont {Merkowitz}}, \bibinfo {author}
  {\bibfnamefont {A.}~\bibnamefont {Morse}}, \bibinfo {author} {\bibfnamefont
  {B.}~\bibnamefont {Price}}, \ and\ \bibinfo {author} {\bibfnamefont
  {N.}~\bibnamefont {Solomonson}},\ }\href {\doibase 10.1103/PhysRevD.54.1264}
  {\bibfield  {journal} {\bibinfo  {journal} {Phys. Rev. D}\ }\textbf {\bibinfo
  {volume} {54}},\ \bibinfo {pages} {1264} (\bibinfo {year}
  {1996})}\BibitemShut {NoStop}%
\bibitem [{\citenamefont {Johnson}\ and\ \citenamefont
  {Merkowitz}(1993)}]{WWJ93}%
  \BibitemOpen
  \bibfield  {author} {\bibinfo {author} {\bibfnamefont {W.~W.}\ \bibnamefont
  {Johnson}}\ and\ \bibinfo {author} {\bibfnamefont {S.~M.}\ \bibnamefont
  {Merkowitz}},\ }\href {\doibase 10.1103/PhysRevLett.70.2367} {\bibfield
  {journal} {\bibinfo  {journal} {Phys. Rev. Lett.}\ }\textbf {\bibinfo
  {volume} {70}},\ \bibinfo {pages} {2367} (\bibinfo {year}
  {1993})}\BibitemShut {NoStop}%
\bibitem [{\citenamefont {Harry}\ \emph {et~al.}(1996)\citenamefont {Harry},
  \citenamefont {Stevenson},\ and\ \citenamefont {Paik}}]{Harry1996}%
  \BibitemOpen
  \bibfield  {author} {\bibinfo {author} {\bibfnamefont {G.~M.}\ \bibnamefont
  {Harry}}, \bibinfo {author} {\bibfnamefont {T.~R.}\ \bibnamefont
  {Stevenson}}, \ and\ \bibinfo {author} {\bibfnamefont {H.~J.}\ \bibnamefont
  {Paik}},\ }\href {\doibase 10.1103/PhysRevD.54.2409} {\bibfield  {journal}
  {\bibinfo  {journal} {Phys. Rev. D}\ }\textbf {\bibinfo {volume} {54}},\
  \bibinfo {pages} {2409} (\bibinfo {year} {1996})}\BibitemShut {NoStop}%
\bibitem [{\citenamefont {Baggio}\ \emph {et~al.}(2005)\citenamefont {Baggio},
  \citenamefont {Bignotto}, \citenamefont {Bonaldi}, \citenamefont {Cerdonio},
  \citenamefont {Conti}, \citenamefont {Falferi}, \citenamefont {Liguori},
  \citenamefont {Marin}, \citenamefont {Mezzena}, \citenamefont {Ortolan},
  \citenamefont {Poggi}, \citenamefont {Prodi}, \citenamefont {Salemi},
  \citenamefont {Soranzo}, \citenamefont {Taffarello}, \citenamefont
  {Vedovato}, \citenamefont {Vinante}, \citenamefont {Vitale},\ and\
  \citenamefont {Zendri}}]{Baggio2005}%
  \BibitemOpen
  \bibfield  {author} {\bibinfo {author} {\bibfnamefont {L.}~\bibnamefont
  {Baggio}}, \bibinfo {author} {\bibfnamefont {M.}~\bibnamefont {Bignotto}},
  \bibinfo {author} {\bibfnamefont {M.}~\bibnamefont {Bonaldi}}, \bibinfo
  {author} {\bibfnamefont {M.}~\bibnamefont {Cerdonio}}, \bibinfo {author}
  {\bibfnamefont {L.}~\bibnamefont {Conti}}, \bibinfo {author} {\bibfnamefont
  {P.}~\bibnamefont {Falferi}}, \bibinfo {author} {\bibfnamefont
  {N.}~\bibnamefont {Liguori}}, \bibinfo {author} {\bibfnamefont
  {A.}~\bibnamefont {Marin}}, \bibinfo {author} {\bibfnamefont
  {R.}~\bibnamefont {Mezzena}}, \bibinfo {author} {\bibfnamefont
  {A.}~\bibnamefont {Ortolan}}, \bibinfo {author} {\bibfnamefont
  {S.}~\bibnamefont {Poggi}}, \bibinfo {author} {\bibfnamefont {G.~A.}\
  \bibnamefont {Prodi}}, \bibinfo {author} {\bibfnamefont {F.}~\bibnamefont
  {Salemi}}, \bibinfo {author} {\bibfnamefont {G.}~\bibnamefont {Soranzo}},
  \bibinfo {author} {\bibfnamefont {L.}~\bibnamefont {Taffarello}}, \bibinfo
  {author} {\bibfnamefont {G.}~\bibnamefont {Vedovato}}, \bibinfo {author}
  {\bibfnamefont {A.}~\bibnamefont {Vinante}}, \bibinfo {author} {\bibfnamefont
  {S.}~\bibnamefont {Vitale}}, \ and\ \bibinfo {author} {\bibfnamefont {J.~P.}\
  \bibnamefont {Zendri}},\ }\href {\doibase 10.1103/PhysRevLett.94.241101}
  {\bibfield  {journal} {\bibinfo  {journal} {Phys. Rev. Lett.}\ }\textbf
  {\bibinfo {volume} {94}},\ \bibinfo {pages} {241101} (\bibinfo {year}
  {2005})}\BibitemShut {NoStop}%
\bibitem [{\citenamefont {Aguiar}\ \emph {et~al.}(2008)\citenamefont {Aguiar},
  \citenamefont {Andrade}, \citenamefont {Barroso}, \citenamefont {Castro},
  \citenamefont {Costa}, \citenamefont {de~Souza}, \citenamefont {de~Waard},
  \citenamefont {Fauth}, \citenamefont {Frajuca}, \citenamefont {Frossati},
  \citenamefont {Furtado}, \citenamefont {Gratens}, \citenamefont {Maffei},
  \citenamefont {Magalh{\~{a}}es}, \citenamefont {Marinho}, \citenamefont
  {Oliveira}, \citenamefont {Pimentel}, \citenamefont {Remy}, \citenamefont
  {Tobar}, \citenamefont {Abdalla}, \citenamefont {Alves}, \citenamefont
  {Bessada}, \citenamefont {Bortoli}, \citenamefont {Brand{\~{a}}o},
  \citenamefont {Costa}, \citenamefont {de~Ara{\'{u}}jo}, \citenamefont
  {de~Araujo}, \citenamefont {de~Gouveia Dal~Pino}, \citenamefont {de~Paula},
  \citenamefont {de~Rey~Neto}, \citenamefont {Evangelista}, \citenamefont
  {Lenzi}, \citenamefont {Marranghello}, \citenamefont {Miranda}, \citenamefont
  {Oliveira}, \citenamefont {Opher}, \citenamefont {Pereira}, \citenamefont
  {Stellati},\ and\ \citenamefont {Weber}}]{Aguiar_2008}%
  \BibitemOpen
  \bibfield  {author} {\bibinfo {author} {\bibfnamefont {O.~D.}\ \bibnamefont
  {Aguiar}}, \bibinfo {author} {\bibfnamefont {L.~A.}\ \bibnamefont {Andrade}},
  \bibinfo {author} {\bibfnamefont {J.~J.}\ \bibnamefont {Barroso}}, \bibinfo
  {author} {\bibfnamefont {P.~J.}\ \bibnamefont {Castro}}, \bibinfo {author}
  {\bibfnamefont {C.~A.}\ \bibnamefont {Costa}}, \bibinfo {author}
  {\bibfnamefont {S.~T.}\ \bibnamefont {de~Souza}}, \bibinfo {author}
  {\bibfnamefont {A.}~\bibnamefont {de~Waard}}, \bibinfo {author}
  {\bibfnamefont {A.~C.}\ \bibnamefont {Fauth}}, \bibinfo {author}
  {\bibfnamefont {C.}~\bibnamefont {Frajuca}}, \bibinfo {author} {\bibfnamefont
  {G.}~\bibnamefont {Frossati}}, \bibinfo {author} {\bibfnamefont {S.~R.}\
  \bibnamefont {Furtado}}, \bibinfo {author} {\bibfnamefont {X.}~\bibnamefont
  {Gratens}}, \bibinfo {author} {\bibfnamefont {T.~M.~A.}\ \bibnamefont
  {Maffei}}, \bibinfo {author} {\bibfnamefont {N.~S.}\ \bibnamefont
  {Magalh{\~{a}}es}}, \bibinfo {author} {\bibfnamefont {R.~M.}\ \bibnamefont
  {Marinho}}, \bibinfo {author} {\bibfnamefont {N.~F.}\ \bibnamefont
  {Oliveira}}, \bibinfo {author} {\bibfnamefont {G.~L.}\ \bibnamefont
  {Pimentel}}, \bibinfo {author} {\bibfnamefont {M.~A.}\ \bibnamefont {Remy}},
  \bibinfo {author} {\bibfnamefont {M.~E.}\ \bibnamefont {Tobar}}, \bibinfo
  {author} {\bibfnamefont {E.}~\bibnamefont {Abdalla}}, \bibinfo {author}
  {\bibfnamefont {M.~E.~S.}\ \bibnamefont {Alves}}, \bibinfo {author}
  {\bibfnamefont {D.~F.~A.}\ \bibnamefont {Bessada}}, \bibinfo {author}
  {\bibfnamefont {F.~S.}\ \bibnamefont {Bortoli}}, \bibinfo {author}
  {\bibfnamefont {C.~S.~S.}\ \bibnamefont {Brand{\~{a}}o}}, \bibinfo {author}
  {\bibfnamefont {K.~M.~F.}\ \bibnamefont {Costa}}, \bibinfo {author}
  {\bibfnamefont {H.~A.~B.}\ \bibnamefont {de~Ara{\'{u}}jo}}, \bibinfo {author}
  {\bibfnamefont {J.~C.~N.}\ \bibnamefont {de~Araujo}}, \bibinfo {author}
  {\bibfnamefont {E.~M.}\ \bibnamefont {de~Gouveia Dal~Pino}}, \bibinfo
  {author} {\bibfnamefont {W.}~\bibnamefont {de~Paula}}, \bibinfo {author}
  {\bibfnamefont {E.~C.}\ \bibnamefont {de~Rey~Neto}}, \bibinfo {author}
  {\bibfnamefont {E.~F.~D.}\ \bibnamefont {Evangelista}}, \bibinfo {author}
  {\bibfnamefont {C.~H.}\ \bibnamefont {Lenzi}}, \bibinfo {author}
  {\bibfnamefont {G.~F.}\ \bibnamefont {Marranghello}}, \bibinfo {author}
  {\bibfnamefont {O.~D.}\ \bibnamefont {Miranda}}, \bibinfo {author}
  {\bibfnamefont {S.~R.}\ \bibnamefont {Oliveira}}, \bibinfo {author}
  {\bibfnamefont {R.}~\bibnamefont {Opher}}, \bibinfo {author} {\bibfnamefont
  {E.~S.}\ \bibnamefont {Pereira}}, \bibinfo {author} {\bibfnamefont
  {C.}~\bibnamefont {Stellati}}, \ and\ \bibinfo {author} {\bibfnamefont
  {J.}~\bibnamefont {Weber}},\ }\href {\doibase 10.1088/0264-9381/25/11/114042}
  {\bibfield  {journal} {\bibinfo  {journal} {Classical and Quantum Gravity}\
  }\textbf {\bibinfo {volume} {25}},\ \bibinfo {pages} {114042} (\bibinfo
  {year} {2008})}\BibitemShut {NoStop}%
\bibitem [{\citenamefont {Arvanitaki}\ and\ \citenamefont
  {Geraci}(2013)}]{PhysRevLett.110.071105}%
  \BibitemOpen
  \bibfield  {author} {\bibinfo {author} {\bibfnamefont {A.}~\bibnamefont
  {Arvanitaki}}\ and\ \bibinfo {author} {\bibfnamefont {A.~A.}\ \bibnamefont
  {Geraci}},\ }\href {\doibase 10.1103/PhysRevLett.110.071105} {\bibfield
  {journal} {\bibinfo  {journal} {Phys. Rev. Lett.}\ }\textbf {\bibinfo
  {volume} {110}},\ \bibinfo {pages} {071105} (\bibinfo {year}
  {2013})}\BibitemShut {NoStop}%
\bibitem [{\citenamefont {Aggarwal}\ \emph
  {et~al.}(2020{\natexlab{b}})\citenamefont {Aggarwal}, \citenamefont
  {Winstone}, \citenamefont {Teo}, \citenamefont {Baryakhtar}, \citenamefont
  {Larson}, \citenamefont {Kalogera},\ and\ \citenamefont
  {Geraci}}]{Aggarwal2020}%
  \BibitemOpen
  \bibfield  {author} {\bibinfo {author} {\bibfnamefont {N.}~\bibnamefont
  {Aggarwal}}, \bibinfo {author} {\bibfnamefont {G.~P.}\ \bibnamefont
  {Winstone}}, \bibinfo {author} {\bibfnamefont {M.}~\bibnamefont {Teo}},
  \bibinfo {author} {\bibfnamefont {M.}~\bibnamefont {Baryakhtar}}, \bibinfo
  {author} {\bibfnamefont {S.~L.}\ \bibnamefont {Larson}}, \bibinfo {author}
  {\bibfnamefont {V.}~\bibnamefont {Kalogera}}, \ and\ \bibinfo {author}
  {\bibfnamefont {A.~A.}\ \bibnamefont {Geraci}},\ }\href@noop {} {\bibfield
  {journal} {\bibinfo  {journal} {arXiv:2010.13157 [gr-qc]}\ } (\bibinfo {year}
  {2020}{\natexlab{b}})}\BibitemShut {NoStop}%
\bibitem [{\citenamefont {Galliou}\ \emph
  {et~al.}(2013{\natexlab{a}})\citenamefont {Galliou}, \citenamefont
  {Goryachev}, \citenamefont {Bourquin}, \citenamefont {Abbe}, \citenamefont
  {Aubry},\ and\ \citenamefont {Tobar}}]{ScRep}%
  \BibitemOpen
  \bibfield  {author} {\bibinfo {author} {\bibfnamefont {S.}~\bibnamefont
  {Galliou}}, \bibinfo {author} {\bibfnamefont {M.}~\bibnamefont {Goryachev}},
  \bibinfo {author} {\bibfnamefont {R.}~\bibnamefont {Bourquin}}, \bibinfo
  {author} {\bibfnamefont {P.}~\bibnamefont {Abbe}}, \bibinfo {author}
  {\bibfnamefont {J.}~\bibnamefont {Aubry}}, \ and\ \bibinfo {author}
  {\bibfnamefont {M.}~\bibnamefont {Tobar}},\ }\href@noop {} {\bibfield
  {journal} {\bibinfo  {journal} {Nature: Scientific Reports}\ }\textbf
  {\bibinfo {volume} {3}} (\bibinfo {year} {2013}{\natexlab{a}})}\BibitemShut
  {NoStop}%
\bibitem [{\citenamefont {Goryachev}\ \emph {et~al.}(2012)\citenamefont
  {Goryachev}, \citenamefont {Creedon}, \citenamefont {Ivanov}, \citenamefont
  {Galliou}, \citenamefont {Bourquin},\ and\ \citenamefont {Tobar}}]{apl2}%
  \BibitemOpen
  \bibfield  {author} {\bibinfo {author} {\bibfnamefont {M.}~\bibnamefont
  {Goryachev}}, \bibinfo {author} {\bibfnamefont {D.~L.}\ \bibnamefont
  {Creedon}}, \bibinfo {author} {\bibfnamefont {E.~N.}\ \bibnamefont {Ivanov}},
  \bibinfo {author} {\bibfnamefont {S.}~\bibnamefont {Galliou}}, \bibinfo
  {author} {\bibfnamefont {R.}~\bibnamefont {Bourquin}}, \ and\ \bibinfo
  {author} {\bibfnamefont {M.~E.}\ \bibnamefont {Tobar}},\ }\href {\doibase
  http://dx.doi.org/10.1063/1.4729292} {\bibfield  {journal} {\bibinfo
  {journal} {Applied Physics Letters}\ }\textbf {\bibinfo {volume} {100}},\
  \bibinfo {eid} {243504} (\bibinfo {year} {2012})}\BibitemShut {NoStop}%
\bibitem [{\citenamefont {Stevens}\ and\ \citenamefont
  {Tiersten}(1986)}]{Tiers1}%
  \BibitemOpen
  \bibfield  {author} {\bibinfo {author} {\bibfnamefont {D.}~\bibnamefont
  {Stevens}}\ and\ \bibinfo {author} {\bibfnamefont {H.}~\bibnamefont
  {Tiersten}},\ }\href@noop {} {\bibfield  {journal} {\bibinfo  {journal} {J.
  Acoust. Soc. Am.}\ }\textbf {\bibinfo {volume} {79}},\ \bibinfo {pages}
  {1811} (\bibinfo {year} {1986})}\BibitemShut {NoStop}%
\bibitem [{\citenamefont {Besson}(1977)}]{1537081}%
  \BibitemOpen
  \bibfield  {author} {\bibinfo {author} {\bibfnamefont {R.~J.}\ \bibnamefont
  {Besson}},\ }in\ \href {\doibase 10.1109/FREQ.1977.200141} {\emph {\bibinfo
  {booktitle} {31st Annual Symposium on Frequency Control}}}\ (\bibinfo {year}
  {1977})\ pp.\ \bibinfo {pages} {147 -- 152}\BibitemShut {NoStop}%
\bibitem [{pz:(1988)}]{pz:1988zr}%
  \BibitemOpen
  \bibfield  {booktitle} {\emph {\bibinfo {booktitle} {ANSI/IEEE Std
  176-1987}},\ }\href {\doibase 10.1109/IEEESTD.1988.79638} {\bibfield
  {journal} {\bibinfo  {journal} {ANSI/IEEE Std 176-1987}\ ,\ \bibinfo {pages}
  {0{\_}1}} (\bibinfo {year} {1988})}\BibitemShut {NoStop}%
\bibitem [{\citenamefont {Galliou}\ \emph
  {et~al.}(2013{\natexlab{b}})\citenamefont {Galliou}, \citenamefont
  {Goryachev}, \citenamefont {Bourquin}, \citenamefont {Abb{\'e}},
  \citenamefont {Aubry},\ and\ \citenamefont {Tobar}}]{Galliou2013}%
  \BibitemOpen
  \bibfield  {author} {\bibinfo {author} {\bibfnamefont {S.}~\bibnamefont
  {Galliou}}, \bibinfo {author} {\bibfnamefont {M.}~\bibnamefont {Goryachev}},
  \bibinfo {author} {\bibfnamefont {R.}~\bibnamefont {Bourquin}}, \bibinfo
  {author} {\bibfnamefont {P.}~\bibnamefont {Abb{\'e}}}, \bibinfo {author}
  {\bibfnamefont {J.~P.}\ \bibnamefont {Aubry}}, \ and\ \bibinfo {author}
  {\bibfnamefont {M.~E.}\ \bibnamefont {Tobar}},\ }\href {\doibase
  10.1038/srep02132} {\bibfield  {journal} {\bibinfo  {journal} {Scientific
  Reports}\ }\textbf {\bibinfo {volume} {3}},\ \bibinfo {pages} {2132}
  (\bibinfo {year} {2013}{\natexlab{b}})}\BibitemShut {NoStop}%
\bibitem [{\citenamefont {Goryachev}\ \emph
  {et~al.}(2014{\natexlab{a}})\citenamefont {Goryachev}, \citenamefont
  {Ivanov}, \citenamefont {van Kann}, \citenamefont {Galliou},\ and\
  \citenamefont {Tobar}}]{ourSQUID}%
  \BibitemOpen
  \bibfield  {author} {\bibinfo {author} {\bibfnamefont {M.}~\bibnamefont
  {Goryachev}}, \bibinfo {author} {\bibfnamefont {E.~N.}\ \bibnamefont
  {Ivanov}}, \bibinfo {author} {\bibfnamefont {F.}~\bibnamefont {van Kann}},
  \bibinfo {author} {\bibfnamefont {S.}~\bibnamefont {Galliou}}, \ and\
  \bibinfo {author} {\bibfnamefont {M.~E.}\ \bibnamefont {Tobar}},\ }\href
  {http://scitation.aip.org/content/aip/journal/apl/105/15/10.1063/1.4898813}
  {\bibfield  {journal} {\bibinfo  {journal} {Applied Physics Letters}\
  }\textbf {\bibinfo {volume} {105}},\  (\bibinfo {year}
  {2014}{\natexlab{a}})}\BibitemShut {NoStop}%
\bibitem [{\citenamefont {Australia}()}]{quake}%
  \BibitemOpen
  \bibfield  {author} {\bibinfo {author} {\bibfnamefont {G.}~\bibnamefont
  {Australia}},\ }\href@noop {} {\enquote {\bibinfo {title}
  {https://earthquakes.ga.gov.au},}\ }\BibitemShut {NoStop}%
\bibitem [{\citenamefont {https://gracedb.ligo.org/latest/}()}]{GraceDb}%
  \BibitemOpen
  \bibfield  {author} {\bibinfo {author} {\bibnamefont
  {https://gracedb.ligo.org/latest/}},\ }\href@noop {} {\enquote {\bibinfo
  {title} {Gracedb},}\ }\BibitemShut {NoStop}%
\bibitem [{\citenamefont {list}()}]{FRBlist}%
  \BibitemOpen
  \bibfield  {author} {\bibinfo {author} {\bibfnamefont {F.}~\bibnamefont
  {list}},\ }\href@noop {} {\enquote {\bibinfo {title} {http://frbcat.org},}\
  }\BibitemShut {NoStop}%
\bibitem [{\citenamefont {Goryachev}\ \emph {et~al.}(2018)\citenamefont
  {Goryachev}, \citenamefont {Kuang}, \citenamefont {Ivanov}, \citenamefont
  {Haslinger}, \citenamefont {M{\"u}ller},\ and\ \citenamefont
  {Tobar}}]{Goryachev:2018aa}%
  \BibitemOpen
  \bibfield  {author} {\bibinfo {author} {\bibfnamefont {M.}~\bibnamefont
  {Goryachev}}, \bibinfo {author} {\bibfnamefont {Z.}~\bibnamefont {Kuang}},
  \bibinfo {author} {\bibfnamefont {E.~N.}\ \bibnamefont {Ivanov}}, \bibinfo
  {author} {\bibfnamefont {P.}~\bibnamefont {Haslinger}}, \bibinfo {author}
  {\bibfnamefont {H.}~\bibnamefont {M{\"u}ller}}, \ and\ \bibinfo {author}
  {\bibfnamefont {M.~E.}\ \bibnamefont {Tobar}},\ }\bibfield  {booktitle}
  {\emph {\bibinfo {booktitle} {IEEE Transactions on Ultrasonics,
  Ferroelectrics, and Frequency Control}},\ }\href {\doibase
  10.1109/TUFFC.2018.2824845} {\bibfield  {journal} {\bibinfo  {journal} {IEEE
  Transactions on Ultrasonics, Ferroelectrics, and Frequency Control}\ }\textbf
  {\bibinfo {volume} {65}},\ \bibinfo {pages} {991} (\bibinfo {year}
  {2018})}\BibitemShut {NoStop}%
\bibitem [{\citenamefont {Lo}\ \emph {et~al.}(2016)\citenamefont {Lo},
  \citenamefont {Haslinger}, \citenamefont {Mizrachi}, \citenamefont
  {Anderegg}, \citenamefont {M\"uller}, \citenamefont {Hohensee}, \citenamefont
  {Goryachev},\ and\ \citenamefont {Tobar}}]{Lo2016}%
  \BibitemOpen
  \bibfield  {author} {\bibinfo {author} {\bibfnamefont {A.}~\bibnamefont
  {Lo}}, \bibinfo {author} {\bibfnamefont {P.}~\bibnamefont {Haslinger}},
  \bibinfo {author} {\bibfnamefont {E.}~\bibnamefont {Mizrachi}}, \bibinfo
  {author} {\bibfnamefont {L.}~\bibnamefont {Anderegg}}, \bibinfo {author}
  {\bibfnamefont {H.}~\bibnamefont {M\"uller}}, \bibinfo {author}
  {\bibfnamefont {M.}~\bibnamefont {Hohensee}}, \bibinfo {author}
  {\bibfnamefont {M.}~\bibnamefont {Goryachev}}, \ and\ \bibinfo {author}
  {\bibfnamefont {M.~E.}\ \bibnamefont {Tobar}},\ }\href {\doibase
  10.1103/PhysRevX.6.011018} {\bibfield  {journal} {\bibinfo  {journal} {Phys.
  Rev. X}\ }\textbf {\bibinfo {volume} {6}},\ \bibinfo {pages} {011018}
  (\bibinfo {year} {2016})}\BibitemShut {NoStop}%
\bibitem [{\citenamefont {Goryachev}\ \emph
  {et~al.}(2014{\natexlab{b}})\citenamefont {Goryachev}, \citenamefont {Farr},
  \citenamefont {Galliou},\ and\ \citenamefont {Tobar}}]{Goryachev:2014ac}%
  \BibitemOpen
  \bibfield  {author} {\bibinfo {author} {\bibfnamefont {M.}~\bibnamefont
  {Goryachev}}, \bibinfo {author} {\bibfnamefont {W.~G.}\ \bibnamefont {Farr}},
  \bibinfo {author} {\bibfnamefont {S.}~\bibnamefont {Galliou}}, \ and\
  \bibinfo {author} {\bibfnamefont {M.~E.}\ \bibnamefont {Tobar}},\ }\bibfield
  {booktitle} {\emph {\bibinfo {booktitle} {Applied Physics Letters}},\ }\href
  {\doibase 10.1063/1.4892926} {\bibfield  {journal} {\bibinfo  {journal}
  {Applied Physics Letters}\ }\textbf {\bibinfo {volume} {105}},\ \bibinfo
  {pages} {063501} (\bibinfo {year} {2014}{\natexlab{b}})}\BibitemShut
  {NoStop}%
\bibitem [{\citenamefont {Goryachev}\ \emph {et~al.}(2020)\citenamefont
  {Goryachev}, \citenamefont {Galliou},\ and\ \citenamefont
  {Tobar}}]{Goryachev:2020aa}%
  \BibitemOpen
  \bibfield  {author} {\bibinfo {author} {\bibfnamefont {M.}~\bibnamefont
  {Goryachev}}, \bibinfo {author} {\bibfnamefont {S.}~\bibnamefont {Galliou}},
  \ and\ \bibinfo {author} {\bibfnamefont {M.~E.}\ \bibnamefont {Tobar}},\
  }\href {\doibase 10.1103/PhysRevResearch.2.023035} {\bibfield  {journal}
  {\bibinfo  {journal} {Physical Review Research}\ }\textbf {\bibinfo {volume}
  {2}},\ \bibinfo {pages} {023035} (\bibinfo {year} {2020})}\BibitemShut
  {NoStop}%
\bibitem [{\citenamefont {Friedt}\ \emph {et~al.}(2005)\citenamefont {Friedt},
  \citenamefont {Mavon}, \citenamefont {Ballandras}, \citenamefont
  {Blondeau-Patissier},\ and\ \citenamefont {Fromm}}]{Friedt:2005aa}%
  \BibitemOpen
  \bibfield  {author} {\bibinfo {author} {\bibfnamefont {J.~.}\ \bibnamefont
  {Friedt}}, \bibinfo {author} {\bibfnamefont {C.}~\bibnamefont {Mavon}},
  \bibinfo {author} {\bibfnamefont {S.}~\bibnamefont {Ballandras}}, \bibinfo
  {author} {\bibfnamefont {V.}~\bibnamefont {Blondeau-Patissier}}, \ and\
  \bibinfo {author} {\bibfnamefont {M.}~\bibnamefont {Fromm}},\ }in\ \href
  {\doibase 10.1109/RADECS.2005.4365630} {\emph {\bibinfo {booktitle} {2005 8th
  European Conference on Radiation and Its Effects on Components and
  Systems}}}\ (\bibinfo {year} {2005})\ pp.\ \bibinfo {pages}
  {PI3--1--PI3--6}\BibitemShut {NoStop}%
\bibitem [{\citenamefont {Lef{\`e}vre}\ \emph {et~al.}(2009)\citenamefont
  {Lef{\`e}vre}, \citenamefont {Devautour-Vinot}, \citenamefont {Cambon},
  \citenamefont {Boy}, \citenamefont {Guibert}, \citenamefont {Chapoulie},
  \citenamefont {Inguimbert}, \citenamefont {Picchedda}, \citenamefont
  {Largeteau}, \citenamefont {Demazeau},\ and\ \citenamefont
  {Cibiel}}]{Lefevre:2009aa}%
  \BibitemOpen
  \bibfield  {author} {\bibinfo {author} {\bibfnamefont {J.}~\bibnamefont
  {Lef{\`e}vre}}, \bibinfo {author} {\bibfnamefont {S.}~\bibnamefont
  {Devautour-Vinot}}, \bibinfo {author} {\bibfnamefont {O.}~\bibnamefont
  {Cambon}}, \bibinfo {author} {\bibfnamefont {J.~J.}\ \bibnamefont {Boy}},
  \bibinfo {author} {\bibfnamefont {P.}~\bibnamefont {Guibert}}, \bibinfo
  {author} {\bibfnamefont {R.}~\bibnamefont {Chapoulie}}, \bibinfo {author}
  {\bibfnamefont {C.}~\bibnamefont {Inguimbert}}, \bibinfo {author}
  {\bibfnamefont {D.}~\bibnamefont {Picchedda}}, \bibinfo {author}
  {\bibfnamefont {A.}~\bibnamefont {Largeteau}}, \bibinfo {author}
  {\bibfnamefont {G.}~\bibnamefont {Demazeau}}, \ and\ \bibinfo {author}
  {\bibfnamefont {G.}~\bibnamefont {Cibiel}},\ }\bibfield  {booktitle} {\emph
  {\bibinfo {booktitle} {Journal of Applied Physics}},\ }\href {\doibase
  10.1063/1.3141750} {\bibfield  {journal} {\bibinfo  {journal} {Journal of
  Applied Physics}\ }\textbf {\bibinfo {volume} {105}},\ \bibinfo {pages}
  {113523} (\bibinfo {year} {2009})}\BibitemShut {NoStop}%
\bibitem [{\citenamefont {Astone}\ \emph {et~al.}(2001)\citenamefont {Astone},
  \citenamefont {Bassan}, \citenamefont {Bonifazi}, \citenamefont {Carelli},
  \citenamefont {Coccia}, \citenamefont {D'Antonio}, \citenamefont {Fafone},
  \citenamefont {Federici}, \citenamefont {Marini}, \citenamefont {Mazzitelli},
  \citenamefont {Minenkov}, \citenamefont {Modena}, \citenamefont {Modestino},
  \citenamefont {Moleti}, \citenamefont {Pallottino}, \citenamefont
  {Pampaloni}, \citenamefont {Pizzella}, \citenamefont {Quintieri},
  \citenamefont {Ronga}, \citenamefont {Terenzi}, \citenamefont {Visco},\ and\
  \citenamefont {Votano}}]{Astone:2001aa}%
  \BibitemOpen
  \bibfield  {author} {\bibinfo {author} {\bibfnamefont {P.}~\bibnamefont
  {Astone}}, \bibinfo {author} {\bibfnamefont {M.}~\bibnamefont {Bassan}},
  \bibinfo {author} {\bibfnamefont {P.}~\bibnamefont {Bonifazi}}, \bibinfo
  {author} {\bibfnamefont {P.}~\bibnamefont {Carelli}}, \bibinfo {author}
  {\bibfnamefont {E.}~\bibnamefont {Coccia}}, \bibinfo {author} {\bibfnamefont
  {S.}~\bibnamefont {D'Antonio}}, \bibinfo {author} {\bibfnamefont
  {V.}~\bibnamefont {Fafone}}, \bibinfo {author} {\bibfnamefont
  {G.}~\bibnamefont {Federici}}, \bibinfo {author} {\bibfnamefont
  {A.}~\bibnamefont {Marini}}, \bibinfo {author} {\bibfnamefont
  {G.}~\bibnamefont {Mazzitelli}}, \bibinfo {author} {\bibfnamefont
  {Y.}~\bibnamefont {Minenkov}}, \bibinfo {author} {\bibfnamefont
  {I.}~\bibnamefont {Modena}}, \bibinfo {author} {\bibfnamefont
  {G.}~\bibnamefont {Modestino}}, \bibinfo {author} {\bibfnamefont
  {A.}~\bibnamefont {Moleti}}, \bibinfo {author} {\bibfnamefont {G.~V.}\
  \bibnamefont {Pallottino}}, \bibinfo {author} {\bibfnamefont
  {V.}~\bibnamefont {Pampaloni}}, \bibinfo {author} {\bibfnamefont
  {G.}~\bibnamefont {Pizzella}}, \bibinfo {author} {\bibfnamefont
  {L.}~\bibnamefont {Quintieri}}, \bibinfo {author} {\bibfnamefont
  {F.}~\bibnamefont {Ronga}}, \bibinfo {author} {\bibfnamefont
  {R.}~\bibnamefont {Terenzi}}, \bibinfo {author} {\bibfnamefont
  {M.}~\bibnamefont {Visco}}, \ and\ \bibinfo {author} {\bibfnamefont
  {L.}~\bibnamefont {Votano}},\ }\href {\doibase
  https://doi.org/10.1016/S0370-2693(01)00026-0} {\bibfield  {journal}
  {\bibinfo  {journal} {Physics Letters B}\ }\textbf {\bibinfo {volume}
  {499}},\ \bibinfo {pages} {16} (\bibinfo {year} {2001})}\BibitemShut
  {NoStop}%
\bibitem [{\citenamefont {Astone}\ \emph {et~al.}(2008)\citenamefont {Astone},
  \citenamefont {Babusci}, \citenamefont {Bassan}, \citenamefont {Bonifazi},
  \citenamefont {Cavallari}, \citenamefont {Coccia}, \citenamefont
  {D’Antonio}, \citenamefont {Fafone}, \citenamefont {Giordano},
  \citenamefont {Ligi}, \citenamefont {Marini}, \citenamefont {Mazzitelli},
  \citenamefont {Minenkov}, \citenamefont {Modena}, \citenamefont {Modestino},
  \citenamefont {Moleti}, \citenamefont {Pallottino}, \citenamefont {Pizzella},
  \citenamefont {Quintieri}, \citenamefont {Rocchi}, \citenamefont {Ronga},
  \citenamefont {Terenzi},\ and\ \citenamefont {Visco}}]{astone2008}%
  \BibitemOpen
  \bibfield  {author} {\bibinfo {author} {\bibfnamefont {P.}~\bibnamefont
  {Astone}}, \bibinfo {author} {\bibfnamefont {D.}~\bibnamefont {Babusci}},
  \bibinfo {author} {\bibfnamefont {M.}~\bibnamefont {Bassan}}, \bibinfo
  {author} {\bibfnamefont {P.}~\bibnamefont {Bonifazi}}, \bibinfo {author}
  {\bibfnamefont {G.}~\bibnamefont {Cavallari}}, \bibinfo {author}
  {\bibfnamefont {E.}~\bibnamefont {Coccia}}, \bibinfo {author} {\bibfnamefont
  {S.}~\bibnamefont {D’Antonio}}, \bibinfo {author} {\bibfnamefont
  {V.}~\bibnamefont {Fafone}}, \bibinfo {author} {\bibfnamefont
  {G.}~\bibnamefont {Giordano}}, \bibinfo {author} {\bibfnamefont
  {C.}~\bibnamefont {Ligi}}, \bibinfo {author} {\bibfnamefont {A.}~\bibnamefont
  {Marini}}, \bibinfo {author} {\bibfnamefont {G.}~\bibnamefont {Mazzitelli}},
  \bibinfo {author} {\bibfnamefont {Y.}~\bibnamefont {Minenkov}}, \bibinfo
  {author} {\bibfnamefont {I.}~\bibnamefont {Modena}}, \bibinfo {author}
  {\bibfnamefont {G.}~\bibnamefont {Modestino}}, \bibinfo {author}
  {\bibfnamefont {A.}~\bibnamefont {Moleti}}, \bibinfo {author} {\bibfnamefont
  {G.}~\bibnamefont {Pallottino}}, \bibinfo {author} {\bibfnamefont
  {G.}~\bibnamefont {Pizzella}}, \bibinfo {author} {\bibfnamefont
  {L.}~\bibnamefont {Quintieri}}, \bibinfo {author} {\bibfnamefont
  {A.}~\bibnamefont {Rocchi}}, \bibinfo {author} {\bibfnamefont
  {F.}~\bibnamefont {Ronga}}, \bibinfo {author} {\bibfnamefont
  {R.}~\bibnamefont {Terenzi}}, \ and\ \bibinfo {author} {\bibfnamefont
  {M.}~\bibnamefont {Visco}},\ }\href {\doibase
  https://doi.org/10.1016/j.astropartphys.2008.09.002} {\bibfield  {journal}
  {\bibinfo  {journal} {Astroparticle Physics}\ }\textbf {\bibinfo {volume}
  {30}},\ \bibinfo {pages} {200} (\bibinfo {year} {2008})}\BibitemShut
  {NoStop}%
\bibitem [{\citenamefont {Spalding}\ \emph {et~al.}(2017)\citenamefont
  {Spalding}, \citenamefont {Tencer}, \citenamefont {Sweatt}, \citenamefont
  {Conley}, \citenamefont {Hogan}, \citenamefont {Boslough}, \citenamefont
  {Gonzales},\ and\ \citenamefont {Spurn{\'y}}}]{Spalding:2017aa}%
  \BibitemOpen
  \bibfield  {author} {\bibinfo {author} {\bibfnamefont {R.}~\bibnamefont
  {Spalding}}, \bibinfo {author} {\bibfnamefont {J.}~\bibnamefont {Tencer}},
  \bibinfo {author} {\bibfnamefont {W.}~\bibnamefont {Sweatt}}, \bibinfo
  {author} {\bibfnamefont {B.}~\bibnamefont {Conley}}, \bibinfo {author}
  {\bibfnamefont {R.}~\bibnamefont {Hogan}}, \bibinfo {author} {\bibfnamefont
  {M.}~\bibnamefont {Boslough}}, \bibinfo {author} {\bibfnamefont
  {G.}~\bibnamefont {Gonzales}}, \ and\ \bibinfo {author} {\bibfnamefont
  {P.}~\bibnamefont {Spurn{\'y}}},\ }\href {\doibase 10.1038/srep41251}
  {\bibfield  {journal} {\bibinfo  {journal} {Scientific Reports}\ }\textbf
  {\bibinfo {volume} {7}},\ \bibinfo {pages} {41251} (\bibinfo {year}
  {2017})}\BibitemShut {NoStop}%
\bibitem [{\citenamefont {Dudorov}\ and\ \citenamefont
  {Eretnova}(2020)}]{Dudorov:2020aa}%
  \BibitemOpen
  \bibfield  {author} {\bibinfo {author} {\bibfnamefont {A.~E.}\ \bibnamefont
  {Dudorov}}\ and\ \bibinfo {author} {\bibfnamefont {O.~V.}\ \bibnamefont
  {Eretnova}},\ }\href {\doibase 10.1134/S003809462003003X} {\bibfield
  {journal} {\bibinfo  {journal} {Solar System Research}\ }\textbf {\bibinfo
  {volume} {54}},\ \bibinfo {pages} {223} (\bibinfo {year} {2020})}\BibitemShut
  {NoStop}%
\bibitem [{\citenamefont {Pospelov}\ \emph {et~al.}(2013)\citenamefont
  {Pospelov}, \citenamefont {Pustelny}, \citenamefont {Ledbetter},
  \citenamefont {Kimball}, \citenamefont {Gawlik},\ and\ \citenamefont
  {Budker}}]{Pospelov:2013aa}%
  \BibitemOpen
  \bibfield  {author} {\bibinfo {author} {\bibfnamefont {M.}~\bibnamefont
  {Pospelov}}, \bibinfo {author} {\bibfnamefont {S.}~\bibnamefont {Pustelny}},
  \bibinfo {author} {\bibfnamefont {M.~P.}\ \bibnamefont {Ledbetter}}, \bibinfo
  {author} {\bibfnamefont {D.~F.~J.}\ \bibnamefont {Kimball}}, \bibinfo
  {author} {\bibfnamefont {W.}~\bibnamefont {Gawlik}}, \ and\ \bibinfo {author}
  {\bibfnamefont {D.}~\bibnamefont {Budker}},\ }\href {\doibase
  10.1103/PhysRevLett.110.021803} {\bibfield  {journal} {\bibinfo  {journal}
  {Physical Review Letters}\ }\textbf {\bibinfo {volume} {110}},\ \bibinfo
  {pages} {021803} (\bibinfo {year} {2013})}\BibitemShut {NoStop}%
\bibitem [{\citenamefont {Jungman}\ \emph {et~al.}(1996)\citenamefont
  {Jungman}, \citenamefont {Kamionkowski},\ and\ \citenamefont
  {Griest}}]{Jungman:1996aa}%
  \BibitemOpen
  \bibfield  {author} {\bibinfo {author} {\bibfnamefont {G.}~\bibnamefont
  {Jungman}}, \bibinfo {author} {\bibfnamefont {M.}~\bibnamefont
  {Kamionkowski}}, \ and\ \bibinfo {author} {\bibfnamefont {K.}~\bibnamefont
  {Griest}},\ }\href {\doibase https://doi.org/10.1016/0370-1573(95)00058-5}
  {\bibfield  {journal} {\bibinfo  {journal} {Physics Reports}\ }\textbf
  {\bibinfo {volume} {267}},\ \bibinfo {pages} {195} (\bibinfo {year}
  {1996})}\BibitemShut {NoStop}%
\bibitem [{\citenamefont {Angloher}\ \emph {et~al.}(2016)\citenamefont
  {Angloher}, \citenamefont {Bento}, \citenamefont {Bucci}, \citenamefont
  {Canonica}, \citenamefont {Defay}, \citenamefont {Erb}, \citenamefont
  {Feilitzsch}, \citenamefont {Ferreiro~Iachellini}, \citenamefont {Gorla},
  \citenamefont {G{\"u}tlein}, \citenamefont {Hauff}, \citenamefont {Jochum},
  \citenamefont {Kiefer}, \citenamefont {Kluck}, \citenamefont {Kraus},
  \citenamefont {Lanfranchi}, \citenamefont {Loebell}, \citenamefont
  {M{\"u}nster}, \citenamefont {Pagliarone}, \citenamefont {Petricca},
  \citenamefont {Potzel}, \citenamefont {Pr{\"o}bst}, \citenamefont {Reindl},
  \citenamefont {Sch{\"a}ffner}, \citenamefont {Schieck}, \citenamefont
  {Sch{\"o}nert}, \citenamefont {Seidel}, \citenamefont {Stodolsky},
  \citenamefont {Strandhagen}, \citenamefont {Strauss}, \citenamefont {Tanzke},
  \citenamefont {Trinh~Thi}, \citenamefont {T{\"u}rko{\u g}lu}, \citenamefont
  {Uffinger}, \citenamefont {Ulrich}, \citenamefont {Usherov}, \citenamefont
  {Wawoczny}, \citenamefont {Willers}, \citenamefont {W{\"u}strich},\ and\
  \citenamefont {Z{\"o}ller}}]{Angloher:2016aa}%
  \BibitemOpen
  \bibfield  {author} {\bibinfo {author} {\bibfnamefont {G.}~\bibnamefont
  {Angloher}}, \bibinfo {author} {\bibfnamefont {A.}~\bibnamefont {Bento}},
  \bibinfo {author} {\bibfnamefont {C.}~\bibnamefont {Bucci}}, \bibinfo
  {author} {\bibfnamefont {L.}~\bibnamefont {Canonica}}, \bibinfo {author}
  {\bibfnamefont {X.}~\bibnamefont {Defay}}, \bibinfo {author} {\bibfnamefont
  {A.}~\bibnamefont {Erb}}, \bibinfo {author} {\bibfnamefont {F.~v.}\
  \bibnamefont {Feilitzsch}}, \bibinfo {author} {\bibfnamefont
  {N.}~\bibnamefont {Ferreiro~Iachellini}}, \bibinfo {author} {\bibfnamefont
  {P.}~\bibnamefont {Gorla}}, \bibinfo {author} {\bibfnamefont
  {A.}~\bibnamefont {G{\"u}tlein}}, \bibinfo {author} {\bibfnamefont
  {D.}~\bibnamefont {Hauff}}, \bibinfo {author} {\bibfnamefont
  {J.}~\bibnamefont {Jochum}}, \bibinfo {author} {\bibfnamefont
  {M.}~\bibnamefont {Kiefer}}, \bibinfo {author} {\bibfnamefont
  {H.}~\bibnamefont {Kluck}}, \bibinfo {author} {\bibfnamefont
  {H.}~\bibnamefont {Kraus}}, \bibinfo {author} {\bibfnamefont {J.~C.}\
  \bibnamefont {Lanfranchi}}, \bibinfo {author} {\bibfnamefont
  {J.}~\bibnamefont {Loebell}}, \bibinfo {author} {\bibfnamefont
  {A.}~\bibnamefont {M{\"u}nster}}, \bibinfo {author} {\bibfnamefont
  {C.}~\bibnamefont {Pagliarone}}, \bibinfo {author} {\bibfnamefont
  {F.}~\bibnamefont {Petricca}}, \bibinfo {author} {\bibfnamefont
  {W.}~\bibnamefont {Potzel}}, \bibinfo {author} {\bibfnamefont
  {F.}~\bibnamefont {Pr{\"o}bst}}, \bibinfo {author} {\bibfnamefont
  {F.}~\bibnamefont {Reindl}}, \bibinfo {author} {\bibfnamefont
  {K.}~\bibnamefont {Sch{\"a}ffner}}, \bibinfo {author} {\bibfnamefont
  {J.}~\bibnamefont {Schieck}}, \bibinfo {author} {\bibfnamefont
  {S.}~\bibnamefont {Sch{\"o}nert}}, \bibinfo {author} {\bibfnamefont
  {W.}~\bibnamefont {Seidel}}, \bibinfo {author} {\bibfnamefont
  {L.}~\bibnamefont {Stodolsky}}, \bibinfo {author} {\bibfnamefont
  {C.}~\bibnamefont {Strandhagen}}, \bibinfo {author} {\bibfnamefont
  {R.}~\bibnamefont {Strauss}}, \bibinfo {author} {\bibfnamefont
  {A.}~\bibnamefont {Tanzke}}, \bibinfo {author} {\bibfnamefont {H.~H.}\
  \bibnamefont {Trinh~Thi}}, \bibinfo {author} {\bibfnamefont {C.}~\bibnamefont
  {T{\"u}rko{\u g}lu}}, \bibinfo {author} {\bibfnamefont {M.}~\bibnamefont
  {Uffinger}}, \bibinfo {author} {\bibfnamefont {A.}~\bibnamefont {Ulrich}},
  \bibinfo {author} {\bibfnamefont {I.}~\bibnamefont {Usherov}}, \bibinfo
  {author} {\bibfnamefont {S.}~\bibnamefont {Wawoczny}}, \bibinfo {author}
  {\bibfnamefont {M.}~\bibnamefont {Willers}}, \bibinfo {author} {\bibfnamefont
  {M.}~\bibnamefont {W{\"u}strich}}, \ and\ \bibinfo {author} {\bibfnamefont
  {A.}~\bibnamefont {Z{\"o}ller}},\ }\href {\doibase
  10.1103/PhysRevLett.117.021303} {\bibfield  {journal} {\bibinfo  {journal}
  {Physical Review Letters}\ }\textbf {\bibinfo {volume} {117}},\ \bibinfo
  {pages} {021303} (\bibinfo {year} {2016})}\BibitemShut {NoStop}%
\bibitem [{\citenamefont {Juillard}(2008)}]{Juillard:2008aa}%
  \BibitemOpen
  \bibfield  {author} {\bibinfo {author} {\bibfnamefont {A.}~\bibnamefont
  {Juillard}},\ }\href {\doibase 10.1007/s10909-008-9742-5} {\bibfield
  {journal} {\bibinfo  {journal} {Journal of Low Temperature Physics}\ }\textbf
  {\bibinfo {volume} {151}},\ \bibinfo {pages} {806} (\bibinfo {year}
  {2008})}\BibitemShut {NoStop}%
\bibitem [{\citenamefont {Ge}\ \emph {et~al.}(2019)\citenamefont {Ge},
  \citenamefont {Lawson},\ and\ \citenamefont {Zhitnitsky}}]{Ge:2019aa}%
  \BibitemOpen
  \bibfield  {author} {\bibinfo {author} {\bibfnamefont {S.}~\bibnamefont
  {Ge}}, \bibinfo {author} {\bibfnamefont {K.}~\bibnamefont {Lawson}}, \ and\
  \bibinfo {author} {\bibfnamefont {A.}~\bibnamefont {Zhitnitsky}},\ }\href
  {\doibase 10.1103/PhysRevD.99.116017} {\bibfield  {journal} {\bibinfo
  {journal} {Physical Review D}\ }\textbf {\bibinfo {volume} {99}},\ \bibinfo
  {pages} {116017} (\bibinfo {year} {2019})}\BibitemShut {NoStop}%
\bibitem [{\citenamefont {Budker}\ \emph {et~al.}(2020)\citenamefont {Budker},
  \citenamefont {Flambaum},\ and\ \citenamefont {Zhitnitsky}}]{nuggets1}%
  \BibitemOpen
  \bibfield  {author} {\bibinfo {author} {\bibfnamefont {D.}~\bibnamefont
  {Budker}}, \bibinfo {author} {\bibfnamefont {V.}~\bibnamefont {Flambaum}}, \
  and\ \bibinfo {author} {\bibfnamefont {A.}~\bibnamefont {Zhitnitsky}},\
  }\href@noop {} {\bibfield  {journal} {\bibinfo  {journal}
  {arXiv:2003.07363v2}\ } (\bibinfo {year} {2020})}\BibitemShut {NoStop}%
\bibitem [{\citenamefont {Raza}\ \emph {et~al.}(2018)\citenamefont {Raza},
  \citenamefont {Van~Waerbeke},\ and\ \citenamefont
  {Zhitnitsky}}]{Raza:2018aa}%
  \BibitemOpen
  \bibfield  {author} {\bibinfo {author} {\bibfnamefont {N.}~\bibnamefont
  {Raza}}, \bibinfo {author} {\bibfnamefont {L.}~\bibnamefont {Van~Waerbeke}},
  \ and\ \bibinfo {author} {\bibfnamefont {A.}~\bibnamefont {Zhitnitsky}},\
  }\href {\doibase 10.1103/PhysRevD.98.103527} {\bibfield  {journal} {\bibinfo
  {journal} {Physical Review D}\ }\textbf {\bibinfo {volume} {98}},\ \bibinfo
  {pages} {103527} (\bibinfo {year} {2018})}\BibitemShut {NoStop}%
\bibitem [{\citenamefont {Zhitnitsky}(2020)}]{Zhitnitsky:2020aa}%
  \BibitemOpen
  \bibfield  {author} {\bibinfo {author} {\bibfnamefont {A.}~\bibnamefont
  {Zhitnitsky}},\ }\href {\doibase 10.1103/PhysRevD.101.083020} {\bibfield
  {journal} {\bibinfo  {journal} {Physical Review D}\ }\textbf {\bibinfo
  {volume} {101}},\ \bibinfo {pages} {083020} (\bibinfo {year}
  {2020})}\BibitemShut {NoStop}%
\end{thebibliography}%


%

\end{document}